\documentclass[letterpaper,aps,preprintnumbers,longbibliography,notitlepage,longbibliography]{revtex4-1}
\usepackage[latin9]{inputenc}
\usepackage{amsmath}
\usepackage{amssymb}
\usepackage{graphicx}
\usepackage{esint}
\usepackage[scr=boondoxo, scrscaled=1.05]{mathalfa}
\usepackage[hidelinks]{hyperref}
\usepackage{subdepth}

\makeatletter

\pdfpageheight\paperheight
\pdfpagewidth\paperwidth

\pdfoutput=1

\usepackage{epsfig}
\usepackage{graphics}

\usepackage{mciteplus}
\mciteErrorOnUnknownfalse

\makeatother

\begin{document}

\preprint{FERMILAB-PUB-17-481-A}

\title{Models of Exotic Interferometer Cross-Correlations in Emergent Space-Time}

\author{Craig  Hogan}
\affiliation{University of Chicago and Fermilab}

\author{Ohkyung Kwon\footnote{E-mail: \href{mailto:o.kwon@kaist.ac.kr}{o.kwon@kaist.ac.kr}}}
\affiliation{Korea Advanced Institute of Science and Technology (KAIST) and University of Chicago}

\begin{abstract}
A Lorentz invariant framework is developed to  model  the  cross spectrum of two interferometers in a space-time that emerges from a Planck scale quantum system with exact causal symmetry and holographic spacelike rotational correlations.   Space-time relationships between world lines are generated by entanglement of  geometrical states on causal diamonds.  The entanglement is tied to  a unique observable signature: an exotic imaginary broad band cross spectrum, with a frequency structure determined by the layout of the interferometers.  The models will be  used to interpret data from the  reconfigured Fermilab Holometer, and for conceptual design of future experiments.

\end{abstract}
\maketitle

\section*{Introduction}

Einstein's classical theory of dynamical space and time--- general relativity--- is well known to be incompatible with principles of quantum mechanics that govern  other forms of physical energy and information.  This mismatch could be resolved if general relativity itself is actually an approximate description of a quantum system.  
One widely supported   view\,\cite{PhysRevD.27.2885,Jacobson1995,Rovelli2004,Anderson:2010xm,Ashtekar:2012np,Verlinde2011,Padmanabhan:2013nxa,PhysRevA.89.052122,Jacobson:2015hqa} is that    space and time  are  statistical or ``emergent'' behaviors of a quantum system with many degrees of freedom, in the same way that sound waves in a gas are a collective behavior of many  quanta.  
The quantum states of the geometrical system should not themselves resemble particles and waves of the familiar sort, which require a background space and time for their definition.   


The well established nonlocality of quantum states\,\cite{RevModPhys.71.S288}--- referred to by Einstein as ``spooky''---  should also apply to the states of geometry. Exotic  geometrical correlations in position can be delocalized in the same way as the wave functions of widely separated  particle pairs. Such  correlations extend indefinitely on null surfaces, limited only by the causal structure of the background space.  Thus, even if timelike correlations and spacelike displacements are locally confined to some microscopic scale--- perhaps as small as the Planck length, {\fontsize{9.6}{0}$\smash[t]{\ell_P= \sqrt{\smash[b]{\hbar G/ c^3}}= 1.6\times 10^{-35}\,}$}m, where quantum wave functions are  comparable in size to the space-time curvature radius of their gravity--- in a long duration measurement, they should lead to  spacelike correlations over  macroscopic separations\,\cite{Hogan:2015b,Hogan:2016}.  The nature of the correlation depends on how elemental degrees of freedom of the geometrical system project onto the measurement, which is the subject of this paper.

An important clue about large scale correlations of quantum geometry comes from thought experiments based on consistency with thermodynamics  of matter in global solutions with horizons, especially black holes, which indicate that whole physical systems have a much smaller   information content than quantum fields\,\cite{Bekenstein1972,Bekenstein1973,Hawking1975}.
For a field system with a UV cutoff at the Planck scale,  information scales like the volume in Planck units, with approximately one independent degree of freedom per Planck volume, whereas the degrees of freedom of  quantum geometry  are  ``holographic''\,\cite{tHooft1993,Susskind1995,Bousso:1999xy}: the information associated with any timelike interval is one quarter of the bounding area of its causal diamond expressed in units of the Planck length, rather than the volume.   If that is the case,  the {density}  of information decreases linearly with the size of the system.
The geometrical  degrees of freedom are less independent than those of quantum fields, {\it especially at large separations}.

This holographic decrease of information density with size 
{\it requires}   exotic, nonlocal spacelike correlations of quantum geometrical states on all scales.  Indeed, well known black hole information ``paradoxes'' follow if information is falsely assumed to be localized\,\cite{Giddings:2012bm,Nomura:2012cx,Hooft:2016pmw,Hooft:2016itl,Hooft:2016cpw}.  Although exotic correlations are  not present in standard formulations of local quantum field theory, they  are displayed explicitly in other model systems, such as dualities between states in anti-de Sitter space and conformal fields on its boundary (``AdS/CFT'')\,\cite{Maldacena1998,Ryu:2006bv,Ryu:2006ef,Solodukhin:2011gn}.  
They must also be present in flat space-time, because the thermodynamic behavior of black hole horizons also applies to Unruh horizon entropy in accelerated frames.  Thus, the same holographic  information bounds  apply, and exotic nonlocal correlations exist, for causal diamonds of any size in any space-time. They  reflect properties of correlations in the quantum system whose emergent statistical behavior gives rise to the classical Einstein equations.

The  AdS/CFT and  black hole examples  show that the exotic information is not localized, but they do not address in detail how  geometrical correlations  affect measurements within the bulk volume on  scales much smaller than the radius of curvature.  In particular,  it is not known how holographic correlations  would manifest themselves in actual measurements  of nonlocal spacelike-separated positions. To address this issue, as is usual in quantum mechanics,  it is necessary to consider  the detailed context of a particular   measurement.  In this paper, we develop a  framework for describing and analyzing measurements of  nonlocal spacelike correlations in the specific context of interferometry. The framework is based on the principle that,  when interpreted  in terms of concrete observables,  exotic correlations should have  the same physical effect on  light   in flat space-time as they do on light near a black hole event horizon.

Our study is based on a  model  where classical projections of geometrical quantum elements live on null cones and causal diamonds associated with a measurement\,\cite{Hogan:2015b,Hogan:2016}. By construction, the model satisfies Lorentz invariance  and holographic information content, and implies a specific structure of nonlocal spacelike correlation that ultimately reflects a hypothesis about exact symmetries at the Planck scale. 
 Although the new exotic quantum displacements are confined to the Planck scale, exotic spacelike correlations of relative position extend everywhere on past and future light cones,  so they encompass all events within the causal diamond associated with a  measurement. 
Measured  properties of the  classical space-time, such as  measurements of spatial directions  and the local inertial frame,  emerge as approximate properties of this Planck scale system.    
The model does not include space-time curvature (which should also be emergent, as a property of excited states), but it enables calculations of  exotic  correlations on measurements in the  case of a nearly-flat space-time, which has no dynamics.
It  predicts departures from standard physics that are small and subtle enough to escape detection in all laboratory experiments to date, but may be large enough to measure in new kinds of experiments.

\subsection*{The Holometer: a new kind of measurement}

The Holometer program is based on the idea that geometrical quantum states, and hence correlations, affect the propagation of light on macroscopic scales.  It may be possible to measure exotic quantum correlations  of  geometry in a macroscopic laboratory apparatus of sufficient sensitivity, in much the same way that spooky spacelike correlations of particle properties have been  studied on macroscopic scales.  

The most sensitive technique for differential position measurement  is laser interferometry \cite{Adhikari:2014}.
An interferometer signal measures the displacement of mirrors in some arrangement in space. The layout of the light paths determines the coupling of correlated geometrical fluctuations to the measurement of differential optical phase. The approximate magnitude of  correlations can be estimated by straightforward extrapolations of quantum mechanics and relativity \cite{Hogan:2008a, Hogan:2008b,  Hogan:2012}. Concrete predictions require a model for the response of a  particular apparatus to exotic correlations\,\cite{Kwon:2014, Hogan:2015a}.

The Fermilab Holometer\,\cite{Holo:Instrument} is the first instrument capable of measuring  correlations of relative positions in four dimensions with sufficient sensitivity to  constrain models of Planck scale holographic correlations.   It consists of two nearly co-located, but independent and isolated, power-recycled Michelson interferometers, whose outputs are cross-correlated. The cross  correlation of differential position across the apparatus is measured over a broad band of frequencies up to and exceeding the inverse light crossing time in the light path. Correlations at  spacelike  separations, determined by the optical layout of the two interferometers, are measured in the time dimension represented by the signal streams. 
Compared with gravitational wave detectors such as LIGO, VIRGO and GEO600,  the Holometer is much more compact (40 m compared with 4 km) and operates at much higher sampling frequency and signal bandwidth ($>$ 1 MHz compared to $<$ 10 kHz).   Unlike gravitational wave detectors, the Holometer measures geometrical strain at frequencies high compared to its inverse light travel time, so exotic spacelike cross correlations on causal diamonds are recorded as timelike cross correlations in the signal. In the Holometer, the correlation between two separate interferometers with fast sampling also allows an unprecedented control over conventional sources of  noise, so that the  sensitivity to correlations in the detection band is limited primarily by fundamental photon quantum noise.

A measure of timelike correlation  often quoted in the gravitational wave literature is the strain  noise power spectral density of fractional position displacement, $h^2$, which has the dimensions of time. For a cross correlation  between two measured displacements $A$ and $B$, it can be written as \begin{align}
\label{eq:powersd}
h^2(f) &\equiv \int_{-\infty}^{\infty} \left\langle \frac{\delta\hspace{-.1em}L_A(t)}{L} \, \frac{\delta\hspace{-.1em}L_B(t - \tau)}{L} \right\rangle_{\hspace{-.25em}t} \, e^{- 2\pi i \tau f} \, d\tau \, ,
\end{align}
where $\langle\hspace{.1em}\rangle_t$ refers to an average in proper time on  a world line.   
Position fluctuations $\delta L/L$ can be visualized as fractional displacements of objects (such as mirrors) separated  by distance $L$. 
The framework developed below adopts an equivalent  power spectral density,  $\smash{ \tilde{S}(f) = h^2 (L/c)^2 }$, that does not explicitly depend on $L$ for its normalization.

In strain units, the expected scale of exotic correlations, normalized by the information in black hole event horizons on scale $L$, is about a Planck time~\cite{Hogan:2008a, Hogan:2008b,  Hogan:2012}:  {\fontsize{9.6}{0}$\smash[t]{h^2\approx t_P\equiv \sqrt{\smash[b]{\hbar G/ c^5}} = 5.4\times 10^{-44}\,}$}sec.  
In the Holometer, this sensitivity is achieved via averages over many cross spectra, in which the uncorrelated standard quantum noise of the laser light averages to zero. Roughly speaking, the mean phase differences of $N \approx 10^{28}$ photons with frequency $f\approx 10^{15}$ Hz can measure correlations of $h^2  \approx (Nf)^{-1}\approx t_P$. 
Unlike proposed AMO experiments on quantum gravity that require large coherent states\,\cite{Marletto:2017kzi,Bose:2017nin}, we avoid this difficulty by sampling superluminally, faster than decoherence.
The Holometer has now made measurements well beyond the Planck threshold of  sensitivity, and a broad class of response models has  been conclusively ruled out, by real data, to about an order of magnitude beyond the holographic information bound~\cite{holoshear}.

\begin{figure}
\begin{centering}
\includegraphics[height=2in]{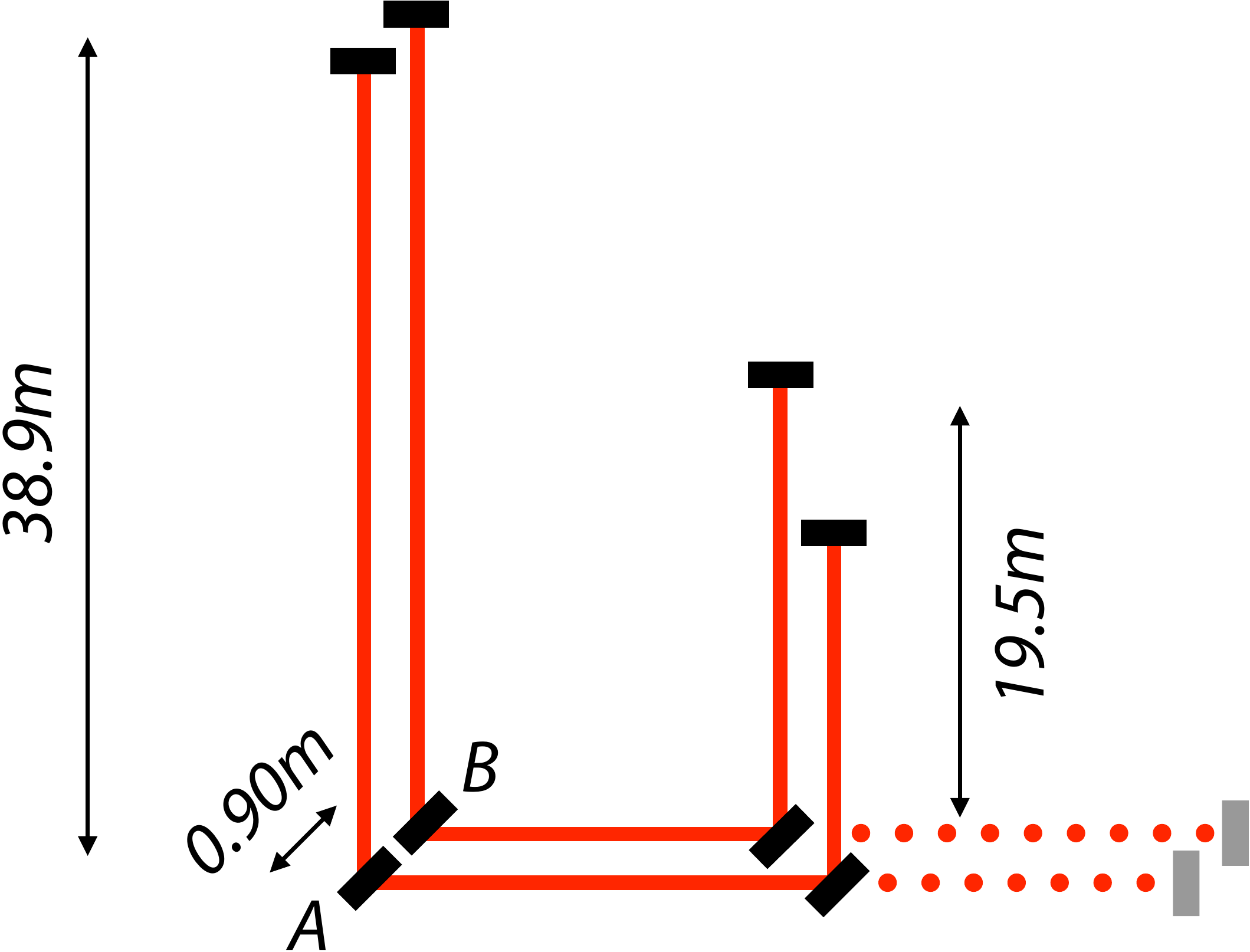}
\par\end{centering}
\protect\caption{Previous (dotted) and reconfigured (solid) paths of light of the two Holometer  interferometers in the lab frame, showing the locations of the mirrors and beamsplitters. Approximate dimensions are shown that enter into  simple models of cross spectrum.\label{layout}}
\end{figure}

\begin{figure}
\begin{centering}
\includegraphics[height=2.75in]{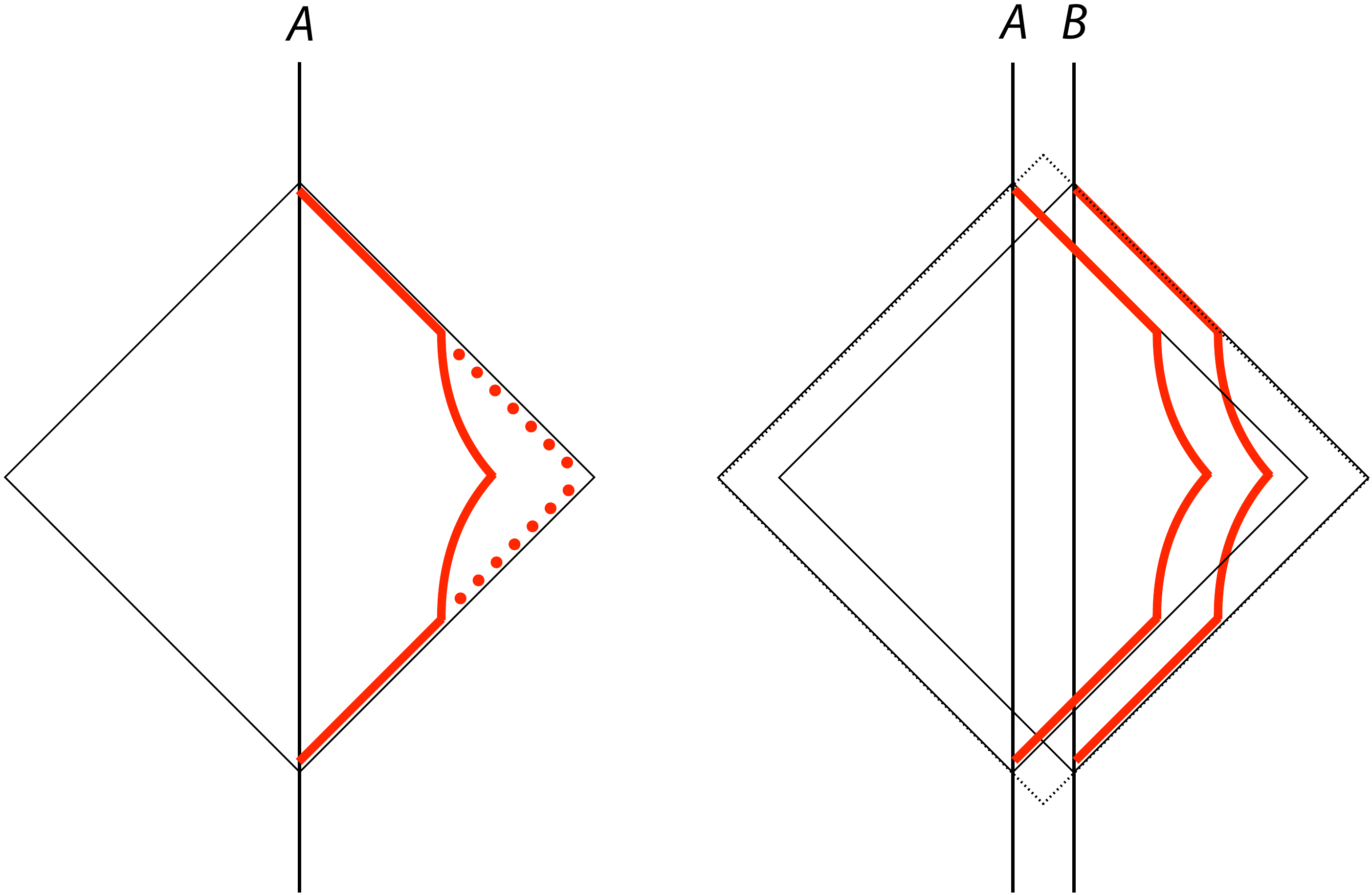}
\par\end{centering}
\protect\caption{Schematic space-time representation of  previous (dotted) and reconfigured (solid) paths of light in the east arms of the Holometer.  Vertical lines are world lines of the beamsplitter(s). On the left, the causal diamond is  shown associated with a free spectral range, a complete round trip of light in the arm.  The radial separation of a ``tracer photon'' in the bent arm is shown in the lab frame; see \cite{Hogan:2016} for a more detailed description.  The transverse part of light path for one arm, where light does not travel radially, departs from the overall causal diamond. The right side shows  two nearby interferometers whose signals are correlated, including the causal diamond associated with a slightly larger minimum interval associated with the cross correlation at zero time lag in the lab frame. \label{newconfiguration}}
\end{figure}

The null result of this experiment represents a constraint on Planck scale physics. It does not rule out the holographic principle, but it constrains the form that holographic correlations can take, and their effect on light.  It is significant that
the null result was obtained with straight interferometer arms, that is, light paths that extend radially from the beamsplitter.
Because of the radial configuration of its light paths, the  Holometer experiment was not able to detect purely transverse correlations, with rotational symmetry around the observer.  
This form of exotic correlation is  better motivated than the translational correlations posited by previous models, since it can be described in a manifestly Lorentz covariant way that accounts for the statistical emergence of local inertial frames in the classical limit \cite{Hogan:2015b, Hogan:2016}.

After obtaining the null result,  the Fermilab Holometer was  reconfigured  (see Figs.~\ref{layout} and \ref{newconfiguration}) to have a nonvanishing response to exotic rotational correlations.  The design was based on  a  response model that  used classical time-of-flight to provide  projections for autocorrelations of one interferometer\,\cite{Hogan:2016}.    
Apart from the addition of new mirrors to bend the light paths, the  new system has not been modified from the one that previously measured a null.  



This paper extends our calculations of  the autospectrum  to model  the actual measured quantity, the cross spectrum between two  interferometers. Motivated by early data in the reconfigured experiment, a new framework is developed  to interpret the relationship between measurements on two  world lines.
Simple models of response in this framework  show  signatures  in the cross spectrum that differ substantially from the previously computed autospectrum.
A series of models is developed, starting with simple illustrations of basic theoretical elements and experimental signatures,  then adding more realistic features to interpret actual data. 



\section*{Basic Principles}


Although the  elemental geometrical degrees of freedom  are not assumed to be conventional particles, waves, or field amplitudes, any measurable correlations must be consistent with basic principles:   the universality and observer-independence of physical laws,  and known symmetries of space-time, particularly Lorentz invariance, in the classical limit.
Observables cannot depend on any coordinate choice of velocity, location or direction, only relative locations and directions determined by the  apparatus.
The framework assumes classical
causal structure and holographic information content, but enlarges the scope of previous models\,\cite{Hogan:2016} to accommodate the effects of quantum entanglement between signals formed in different  interferometers. The  basic principles are: 

\begin{enumerate}
\item
Measurements are observables  in a quantum system whose geometrical states live on  causal diamonds.

Positions of events are not fundamental entities, but observables within a quantum system  that  gives rise to macroscopic positional relationships in space and time.
The system  gives rise to   concepts such as locality, distance, direction, and duration, which rely on  space-time for their definition.
By construction, it exactly preserves causal structure for any measurement.
Classical space-time is ``made of'' localized events, but its quantum elements are ``made of''  nonlocal invariant structures that approximate null cones on large scales (see Fig.\,\ref{foliation}).  

\begin{figure}
\begin{centering}
\includegraphics[height=2.65in]{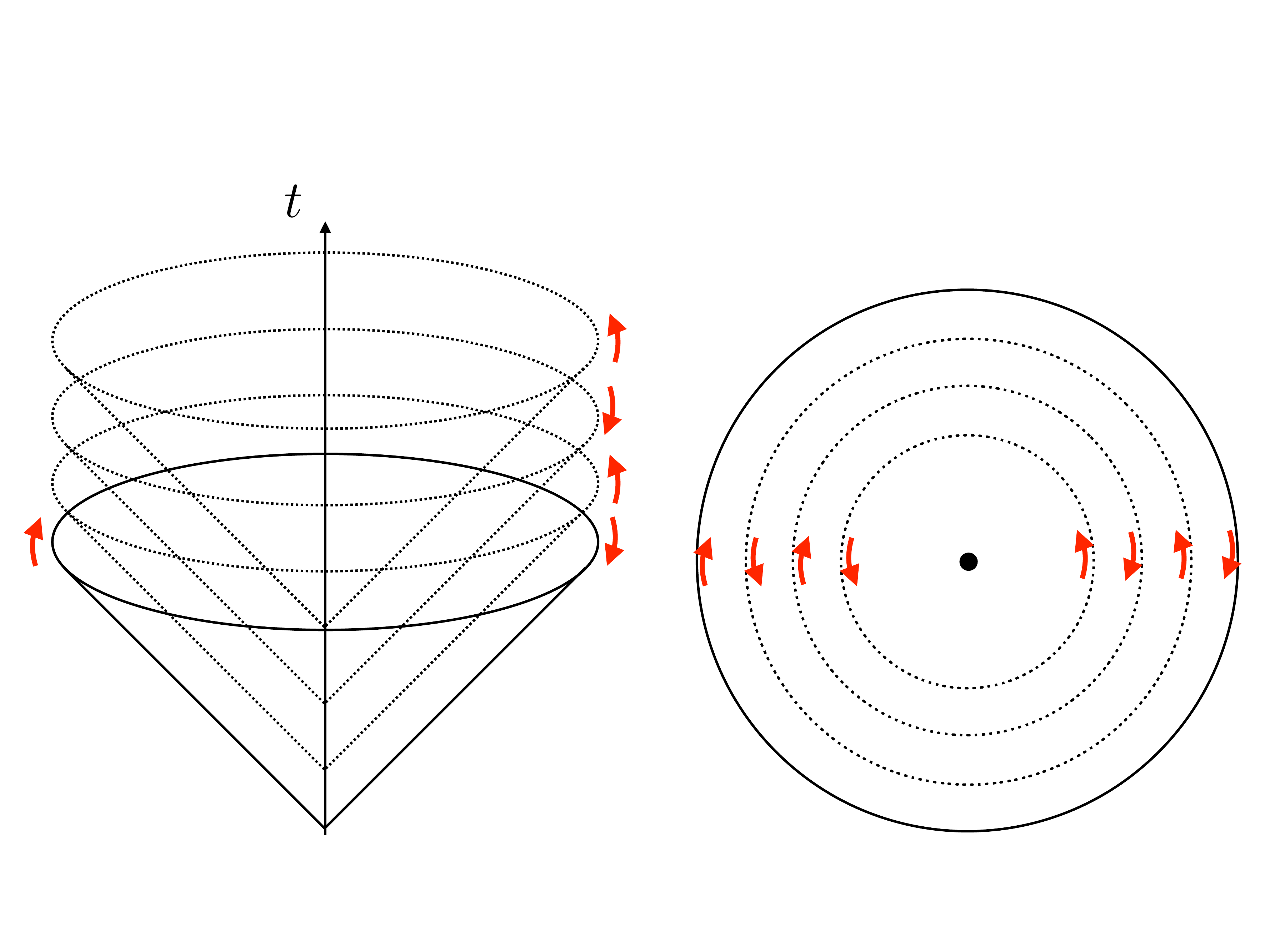}
\par\end{centering}
\protect\caption{Causal structure of exotic correlations in a measurement of quantum geometry. Left, foliation of space-time by light cones of an observer's world line.  Timelike correlations with Planck bandwidth in proper time on the world line are shared everywhere on light cones. Arrows represent correlated transverse displacements (``twists'') at the Planck scale.  Right, projection of multiple light cones onto a surface of constant time in the rest frame. Correlations vanish for timelike separations larger than $t_P$, but spacelike correlations extend indefinitely on light cones. The actual sequence of displacements depends on incoming environmental information that includes positional relationships with other world lines, which are not shown here. \label{foliation}}
\end{figure}

\item
The system is holographic: the number of degrees of freedom in a causal diamond is the same as that of a black hole of the same radius.

The holographic information content  arises from a  Planck bandwidth of information in proper time along any world line, or (equivalently) from a discretization of timelike intervals at the Planck time.  
From each event on the world line, the states of the light cones have Planck scale indeterminacies along the spacelike surfaces of 2-spheres, with correlations for two independent rotational axes. 
The correlations limit  the total information to the bounding surface area of causal diamonds in Planck units.

Similar nonlocal exotic correlations based on covariant causal structure should also appear in excited or thermal systems built of the same quantum elements, such as curved emergent space-times with horizons.
For example, nonlocal correlations in black holes and AdS/CFT spacetimes have been extensively studied in  frameworks based on entanglement entropy\,\cite{Solodukhin:2011gn}.   
 The framework developed here can be thought of as a specific physical implementation of entanglement-entropy correlations in flat space-time.
 The  emergence of gravity is not included because we do not study the effects of any dynamics or emergent curvature, only correlations; in this sense, our framework represents a special-relativity limit of quantum gravity.
The aim here is to compute  effects of the exotic correlations on light that can be measured in actual experiments.
\item
Geometrical correlations on null surfaces affect transversely propagating states of light.

Our approach  connects to gravity by the (controversial) hypothesis that  covariant properties of exotic correlations represent the same geometrical quantum degrees of freedom for any space-time, and thus have an equivalent effect on fields:  {\it  When interpreted  in terms of concrete observables,  they should have  the same physical displacement effect on states of light propagating tangent to a light cone in flat space-time that they do on tangential components of field states near a black hole event horizon}.

\end{enumerate}

The flat-space-time foliation  can then be precisely related to the entropy of a black hole event horizon, as described in the Appendix.  
 When foliated null cone states are mapped onto the outgoing null surfaces near the event horizon of a Schwarzschild black hole, their correlations have a finite angular resolution around each independent axis,  at the angular scale subtended by one Planck length at the Schwarzschild radius. 
The geometrical states thereby entangle incoming and outgoing particle states at the angular wavenumber required to match the geometrical holographic information content of the horizon \cite{Hooft:2016pmw,Hooft:2016itl,Hooft:2016cpw}.   
Apart from this normalization, gravity  and curvature do not enter further into the analysis here; some possible  implications are discussed in the Appendix.

\subsection*{Physical picture of rotational correlations and fluctuations in emergent inertial frames}

In any classical space-time,  there is an exact nonrotating local inertial frame at every event, defined by the metric and measurable locally in any infinitesimal volume. In a flat space-time, the local inertial frame is also a global one.
Our exotic correlations can be visualized as an effect of rotational fluctuations in the inertial frame~\cite{Hogan:2015b}---  a nonlocal quantum mechanical effect that is not possible in classical geometry.
 
In an emergent space-time, even if it is nearly flat on large scales, an inertial frame  must be defined operationally by  a nonlocal measurement. There is no absolute space, only a quantum system of some yet-to-be-defined elements whose operations define spatial and temporal relationships, including locality itself.  A measurement defines a proper time, and the separation between radial and  transverse spatial directions.  

In our system, the  observable is an  interferometer signal.   Measurements are associated with timelike intervals on  world lines and their  causal diamonds. The classical limit of the system is an interval of infinite duration. Nonlocal measurements  of finite duration can record rotational fluctuations with respect to the classical inertial frame. These exotic fluctuations, and their correlations,  do not  correspond to any perturbation of a classical metric.

Rotational fluctuations have a different effect on a cross correlation than they do on an autocorrelation.
The autospectrum is a self-contained measurement that depends only on the causal structure of one interferometer and the entanglement of its geometry with the measured light. 
In previous work, we used a time-of-flight calculation in classical space and time to make a prediction for the posited statistical correlations and their mapping onto signal spectra in time and frequency~\cite{Hogan:2016}. This was possible for the autocorrelation because in that case, there is a single frame, associated with  the world line of the measurement at the beamsplitter. 
The cross spectrum, on the other hand, involves two measurements, in two interferometers whose geometrical entanglement depends on how they are arranged relative to each other. Exotic correlations affect not only the ``motion'' of light and mirrors around each world line, but also the relationship between the two world lines. {\it The two measurement frames are not classically inertial with respect to each other.}

New behavior is possible in a cross spectrum that is not possible in an autospectrum. Similar to how a classical rotation produces opposite transverse displacements in the positions of two bodies (relative to a fixed classical direction), the entanglement manifests as a reciprocity in the fluctuations of two inertial frames, leading to an antisymmetric correlation in the time domain and an imaginary cross spectrum in the frequency domain. The distinctive qualitative differences cannot be captured as small perturbations of the earlier classically derived autospectra. They can however be captured by  semiclassical (non-quantum) statistical models of  effects of spooky entanglement on cross correlations, as presented below.


\section*{Exotic Response, Autocorrelations, and Autospectra}

\subsection*{Statistical model of interferometer signals in emergent space-time}

Our model for the formation of the cross correlated signal builds on  Lorentz invariant  elements of signal formation adapted from our previous model of autocorrelation\,\cite{Hogan:2016}.  It is based on the foliation of space-time by invariant causal structure:   Planck bandwidth information on world lines, and quantum states that live on light cones.

In our model of the apparatus, world lines $A$ and $B$  represent the patches on the two beamsplitter mirrors where interference occurs. The apparatus records and correlates two time series, $s_A(t)$ and $s_B(t)$.  The actual signals are averaged voltages on pairs of photodiodes that represent power emerging from the dark ports of the two patches.  The variables $s_A(t)$, $s_B(t)$ represent the instantaneous differential displacement of light phase arriving from the two interferometer arms of $A$ or $B$ at each time in their respective frames.

The signals depend on a  set of directional relationships in space, defined by the apparatus. The patch of interfering light for each interferometer arises from  two arms that define  paths in space.  
In our model for the cross correlation,  the quantum system includes not only the mirrors and light, but also  the quantum geometry that we aim to measure.   
The  Holometer is designed so  that the signal is sensitive to Planck scale correlations that mix time and space.
The quantum aspects of its geometrical state cannot be described using conventional classical time and position variables, but their exotic correlations can.  
If  the signals of the two interferometers are affected by nonlocal spacelike correlations in the quantum geometry, their signals are not independent, but have spooky  quantum correlations.

Let $\Delta_A(t),\, \Delta_B(t)$ denote idealized fundamental time sequences on  world lines $A,B$ that represent projections of the states of quantum geometry onto the time series of the signals.  As we are constructing a statistical model, not a quantum one, they represent real  random variables, not quantum operators; however, they encode the holographically limited information in quantum geometrical observables that depend on relative position.   In continuous laboratory time, we say that they have a Planck bandwidth, so on each world line their values are correlated within a Planck interval of proper time. They could  be described without loss of information as a discrete time series of numbers.
In the frequency domain, their Fourier transforms  are  spectra $\tilde{\Delta}(f)$  with a flat power spectrum,\vspace{0.1em}
\begin{equation}\label{plancknoise}
|\tilde{\Delta}(f)|^2  \approx   t_P.
\end{equation}

\vspace{0.1em}Thus, a sequence in proper time, represented as a series of numbers with Planck bandwidth, can be attached to past and future null cones  that foliate the entire space-time.  For every event, the  null cones represent incoming and outgoing 2-surfaces that travel at   the speed of light. They represent a projection of the exotic geometrical quantum states, and are the source of the exotic nonlocal holographic correlation (Fig.\,\ref{foliation}).  

Lorentz invariance requires that  the spatial projections of $\Delta_A(t),\, \Delta_B(t)$ be statistically isotropic tensor functions, proportional to the antisymmetric Levi-Civita tensor $\smash{\epsilon^{ijk}}$. The 3-indices $ijk$ refer to directions in space: the spatial directions associated with components of  geometrical states (such as rotation axes and transverse displacements of null cone states), and  null vectors that define classical directions. Any actual observables--- in our case,  the signals themselves---  must  be Lorentz scalars obtained by contractions of indices. Most of this paper concentrates on the time and frequency domains; the spatial dependence of the cross correlations is discussed in more detail in the Appendix. 


For  two world line segments where causal diamonds  share a  4-volume of space-time, the two sequences $\Delta_A(t),\, \Delta_B(t)$ are not independent, but are entangled. By correlating them, it should be possible, if quantum geometry is indeed relational,  to reconstruct their spatial relationship, up to bandwidth limitations.  For example, if they are separated by a  distance $R$, nonzero correlations appear at time intervals exceeding $R/c$ that encode their relative positions.   The Planck bandwidth limitations correspond to information loss or noise, which appear as correlations in the holometer measurement. The relational hypothesis says that the  cross correlations between $\Delta_A(t), \,\Delta_B(t)$ contains all the information there is to know about the relative spatial relationship of the two world lines: \textit{it is encoded as pure }``\textit{entanglement information}.''
A simple quantum system to illustrate this entanglement is described in the Appendix. 

The sequences $\Delta_A(t),\, \Delta_B(t)$ represent the information available where the quantum measurement is made, that is, at the $A$ and $B$ patches. In fact, we do not make Planck bandwidth measurements so we do not care about (and the signals do not measure) information from a region smaller than a laser beam waist size, which is a few millimeters. It does not matter, because we are also only measuring entanglements and correlations in causal diamonds    larger than a meter.  The (exotic) claim is that Planck scale correlations are still detectable at our measured frequencies, determined by the macroscopic separation of the optics.

The autonomous spectrum of each interferometer depends on statistics of a single photon interfering with itself, correlated with a single, classical proper time.    Our model of positional cross correlations depends on an  approximation for independent signal formation that  assumes a separate interference (or ``collapse'' of the photon wavefunction) for the two signals, and a cross correlation between them that depends on the relationship between the two world lines.

Let $W(\tau)$ denote  response functions of an  interferometer signal to the geometrical state vector represented by the Planckian sequence  $\Delta(t)$ on the measurement world line.  It depends on the shape of the interferometer: in general $W_A(\tau), \,W_B(\tau)$ are  not the same, although in the Holometer, they are similar.  The individual response functions of each of the two interferometers to the space-time fluctuations should depend only on the shapes of interferometer light paths, not on their location or orientation in space. 
In a  covariant formulation, the response functions are tensors, as discussed in the Appendix.  Their spatial structure  controls the response, although the spatial  indices are generally suppressed in  the following analysis, which focuses on measurements of time and frequency correlations.  

The signal of each interferometer results from a convolution of its
Planck scale geometrical state vector projection $\Delta(t)$,   and the interferometer response $W$:
\begin{equation}\label{signals}
s_{A} (t)  = \int d\tau \Delta_A(t-\tau) W_A(\tau)\,.
\end{equation}
The exotic signal response--- not present in standard physics--- is expressed here in the form of the convolution of the Planck scale sequence with a dimensionless response function $W$ that has support over a macroscopic time interval $\approx L/c$ determined by  arm lengths $L$ of the apparatus. 
The response spectrum  $\tilde{W}(f)$  typically (depending on detailed projection factors) has values $\approx L/c$  up to $f\approx c/L$. 
In the frequency domain, Eq.\,(\ref{signals}) transforms to a power spectrum
\begin{equation}\label{autopower}
\tilde {S}(f) \equiv \tilde {s}(f) \tilde {s}(f)^* = |\tilde{\Delta} (f)|^2 \,|\tilde{W}(f)|^2\,,
\end{equation}
where $*$ denotes conjugation.


\begin{figure}
\begin{centering}
\includegraphics[height=3.55in]{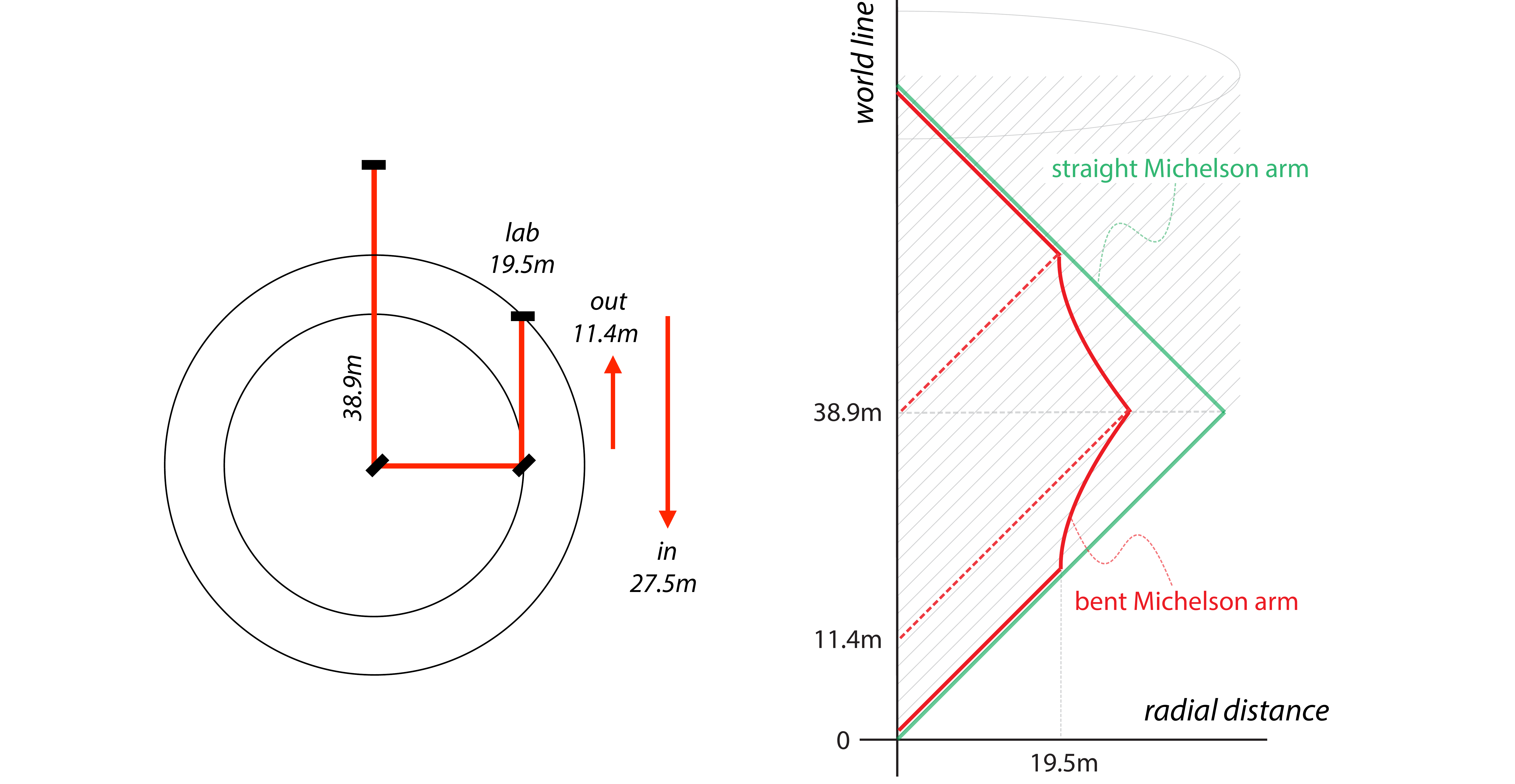}
\par\end{centering}
\protect\caption{Intersections of null cone surfaces at different lab times with the transverse arm, shown in the plane of the interferometer (left) and on a space-time diagram (right). Unlike the toy model, the eventual duration of light affected by this impulse is different for outwards and inwards propagating light, as shown by the arrows.  These segments map onto time and frequency scales of the measured spectrum.\label{impulse}}
\end{figure}

For a rough estimate of exotic noise in the models proposed here,  visualize   a random rotational ``twist,'' or  transverse displacement by one Planck length, by a new null cone every Planck time.  As a result of its encounter  with light in a transverse arm of macroscopic length $L$ (see Figs.~\ref{foliation} and \ref{impulse}),  the light in the arm  eventually appears with a correlated signal displacement on the beamsplitter world line over a time $\approx L/c$. Although each displacement is only about a Planck length, an accumulation of such  coherent displacements leads to signal displacements on scale $L/c$ with
variance   $\smash{\approx (ct_P)^2 (L/ct_P)}$, much larger than a Planck length---   of the order of a Planck random walk.  In spectral language,  the power spectrum $\smash{\tilde {S}}$ is of order $\smash{t_P (L/c)^{2}}$   over bandwidth $\approx c/L$, which leads to a variance $\approx  t_P L/c$. The same scale can be derived in a wave theory of the geometrical states, in the form of a diffraction limit on directional resolution at null separations \cite{Hogan:2008a, Hogan:2008b,  Hogan:2012}.

\subsection*{Boxcar autoresponse  model}


The response function $W$ can be thought of as a  response to a Planck scale impulse or ``twist''--- a transverse Planck scale displacement everywhere on a light cone.  The exotic response differs qualitatively from local correlators of  field theory because the apparatus responds to nonlocal correlations at macroscopic separations on the same light cone.  
Before constructing  realistic models of the response, it is useful to illustrate the basic mechanism and approximate scales with a simple  toy model that shows how transverse components of light path segments convert spacelike correlations on null cones into  timelike macroscopic correlations of a signal.

Consider  the response of  one bent arm that includes a  transverse segment of length $L_\perp$. Suppose the transverse arm segment is  tangent to a null cone from the  world line of the measurement.  Light in the arm is displaced along the transverse arm by a Planck length at  spacelike separated events where the arm intersects the null cone.    At each event, the impulse imposes the same longitudinal displacement on light  traveling in the arm.  The displacement  leads to an eventual signal response--- a displacement in  phase--- when the light arrives  at the beamsplitter from the event where it was displaced, and its phase is measured relative to another arm (see Fig.\,\ref{impulse}).  
 
 In an idealized one-way-time-of-flight ``boxcar'' toy model, the response  is uniformly distributed in laboratory time for light propagating along the arm in one direction.  The response  has  a duration $L_\perp/c$ and unit amplitude: 
 \begin{equation}\label{boxcar}
W(\tau) =\begin{cases}
1\;, & 0< \tau <L_\perp / c\\
0\;, & {\rm otherwise}\, ,
\end{cases}
\end{equation}
where  $\tau=0$ is set  to be the laboratory time lag when the ``first''  light  arrives back at the beamsplitter. The autocorrelation of a boxcar is a triangle, and the frequency power spectrum of such a model is
\begin{equation}\label{toypower}
\tilde{W}\tilde{W}^* =  2\,(L_\perp/c)^2 \, {\rm sinc}^2   (\pi f L_\perp/c)\,.     
\end{equation}
The spectrum has  zeros at integer multiples of $f= c/L_\perp$, determined by the length of the transverse arm. 
Note that Eq.\,(\ref{toypower}) is written in the engineering convention with a prefactor of 2, folding the power at negative frequencies into the positive domain.


\subsection*{Computed time-of-flight autoresponse functions with geometrical projection factors}
A more realistic model of response\,\cite{Hogan:2016}  includes  standard geometrical projection factors in both spatial angle  and  measured laboratory time.
The idealized boxcar response can be conceptually understood as the case of a transverse arm following a circular arc, with a null cone impulse impacting the whole $L_\perp$ segment at the same lab time. In a real experiment, the light paths are composed of straight arm segments between mirrors. Within each segment, a single null cone from a beamsplitter event  intersects a macroscopic segment of propagating light (as in the toy model), but the interval over which the affected light returns in the measured signal is different for inbound and outbound light (see Fig.\,\ref{impulse}), because light in a straight transverse arm segment has a nonzero radial velocity, either parallel or antiparallel to that of the null cone displacement. The inbound and outbound segments correspond to different durations of proper time when projected along null cones to and from the observer world line.

To illustrate the nature of the projection, consider a simple layout similar to the Holometer, an arm of length $2L = 38.9$m  bent at a right angle in the middle so the radial and transverse segments both have length $L=19.5$m.  A pulse arrives at the end of the transverse arm later than it arrives at the bend, by $ L (\sqrt{2} - 1)$.  The detected correlated segment of outgoing radiation affected by the pulse is shortened by this amount, and the incoming radiation is lengthened by the same amount.   The outbound light segment for a single null cone has  a detected duration $\smash{c\Delta \tau = [1- (\sqrt{2} - 1)]L = 11.4}$m, and the  inbound detected segment  has a detected duration $\smash{c\Delta\tau =  [1 + (\sqrt{2} - 1)]L = 27.5}$m  (see Fig.\,\ref{impulse}). Instead of one scale $L_\perp$ as in the single-boxcar model (Eq.\,\ref{boxcar}), a more physical  model of response in the transverse arm has at least two scales, one for each direction of light propagation, $L_{\perp,out} =  11.4$m and $L_{\perp,in}=  27.5$m. 

These time interval segments in $W(\tau)$ roughly determine the characteristic frequencies for the response function $\smash{\tilde{W}(f)}$. Finer structures are also affected by geometrical projection factors between trajectories of light in the arms and null cones or causal diamonds of space-time centered on the beamsplitter. Near the bend mirrors, the light trajectories are purely angular and couple optimally to impulses tangent along the null cones; but closer to the end mirrors, the light paths undergo a mixture of angular and radial propagation, partially suppressing the device's response (see Fig.\,\ref{impulse}).

\begin{figure}
\begin{centering}
\includegraphics[width=7in]{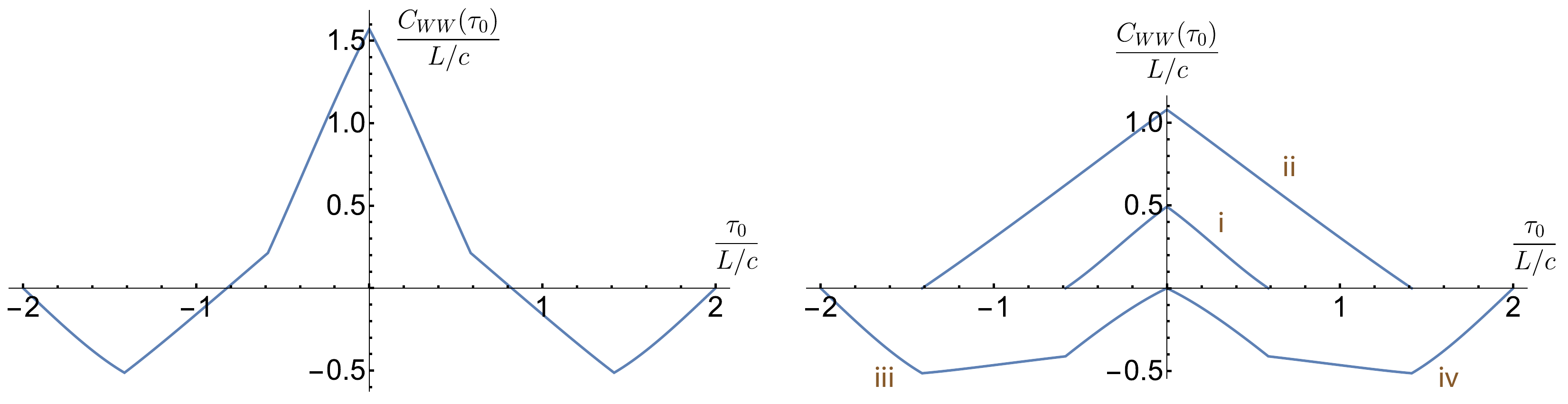}
\par\end{centering}
\protect\caption{The time-domain exotic autocorrelation power for the $W(\tau)$ functions calculated in ref.\,\cite{Hogan:2016}, plotted against time lag $\tau_0$ (left), in units of $L=19.5$m. The same plot is also shown decomposed into components (right). The two triangles (i and ii) can each be approximated as autocorrelations of boxcars of $L_{\perp,out} =  0.59L $ and $L_{\perp,in}=  1.41L$, with the maximum variance respectively reduced to 84.1\% and 76.2\% of unity-slope triangles due to angular projection factors. The two quadrilaterals (iii and iv) can be understood as correlations between boxcars of $L_{\perp,out}$ and $L_{\perp,in}$. They are negative because light travels in opposite directions in these segments. An idealized toy model would generate isosceles trapezoids from correlating two boxcars of different lengths, but angular projection factors lead to some asymmetry. The average power in the slanted base is 46.4\% of the maximum variance that would be reached by a triangle with $L_{\perp,out}+L_{\perp,in}=2L$ total support and unity slope.\label{autodecomposition}}
\end{figure}

These known properties of the interferometer segments can be used to model the response more realistically than the toy boxcar model (Eqs.~\ref{boxcar} and \ref{toypower}). This was done in ref.\,\cite{Hogan:2016}, with the resulting time-domain spectrum shown in Fig.\,\ref{autodecomposition}. It  has a simple intuitive  relationship to the heuristic toy boxcar model: The total time-domain autocorrelation spectrum breaks down into two approximate triangles, each corresponding to autocorrelations of boxcars of $L_{\perp,out}$ and $L_{\perp,in}$, and two approximate quadrilaterals that resemble slanted trapezoids, corresponding to correlations between different-length segments of $L_{\perp,out}$ and $L_{\perp,in}$. Slight asymmetries, nonlinearities, and reductions in power are visible due to angular projection factors. In particular, the maximum variance for the $L_{\perp,out}$ and $L_{\perp,in}$ triangles are respectively reduced to 84.1\% and 76.2\% of autocorrelations from unit amplitude boxcars, and the negative correlation between opposite-direction segments has a broad slanted base with an average power amounting to 46.4\% of the maximum variance in unity-slope triangles of $L_{\perp,out}+L_{\perp,in}=2L$ total support.

\section*{Exotic Cross-Correlations}

Although in principle the displacements represented by $s_{A}$ and $s_{B}$ are present in the raw Holometer data for the individual $A$ and $B$ signals, their autospectra are not measured individually with interesting sensitivity. In order to achieve Planck sensitivity, the Holometer needs to cross correlate signals from two interferometers. 
To account  for the  cross correlation measurement, the autospectrum model must be augmented by adding  other concepts not previously considered.  

The exotic cross correlation is affected by the fact that the two world lines are not inertial with respect to each other. Their rest frames and  proper times do not agree, because of the exotic displacements.

In our approximation scheme, projections by measurement of state vectors onto timelike  world lines are represented by $\Delta$, and  the response functions of signals  to those states represented by $W$. Denote by subscripts $A|B$ and $B|A$  the  parts of the  $A$ measurement correlated with $B$, and {\it vice versa}. 
Again suppressing the dependencies on spatial indices (as discussed in the Appendix), the  frequency transform of the cross correlated signal can be written 
\begin{equation}\label{crossspectrum}
\tilde {S}_{BA} (f) \, =\, ( \tilde{\Delta}_{B|A}  \tilde{W}_{B|A})\, ( \tilde{\Delta}_{A|B}  \tilde{W}_{A|B} )^*
\,= \, ( \tilde{\Delta}_{B|A}  \tilde{\Delta}_{A|B}^* ) \ \,  ( \tilde{W}_{B|A}   \tilde{W}_{A|B}^*)\,.
\end{equation}
 Note that in the frequency domain the two signal transforms are conjugate multiplied, instead of simply multiplied as in Eq.\,(\ref{autopower}), because  in the time domain they are cross correlated instead of convolved.
 
 The  $ \smash{\tilde{\Delta}_{B|A}  \tilde{\Delta}_{A|B}^*}$ term represents the time sequence cross correlation. It arises from the entanglement of the  states of the two emergent world lines as they are projected onto classical proper time. Its behavior should be a universal behavior of emergent geometry, and is highly constrained by our hypothesis of exact causal symmetry. 
 
The  $\smash{\tilde{W}_{B|A}  \tilde{W}_{A|B}^*}$ term represents the  cross projection of the instrument response. It describes the entanglement of the light with the emergent geometrical states, including the shape of the light paths in the 3D space of the rest frame of the apparatus as shaped by the world lines of its mirrors. Its properties are also partially constrained by Lorentz invariance and other mathematical constraints, and by the entanglement required for photon states near black hole horizons.

The goal is to figure out the unknown physics of  cross correlations, notated here as the subscripts $A|B$ and $B|A$,
that  represent the projections of spooky correlations onto the signal.
Writing Eq.\,(\ref{crossspectrum})  in this form already makes physical presumptions, because   
the separation into $\Delta$  and $W$  subsystems depends on the world line projection of emergent states. 
The intent of separation is to connect with physical systems we do understand:  one part  only depends on separation of measurements in the rest frame, and the other  part  depends  only on the geometrical layout of the light paths in the two  measurements.
This approximation is not valid at all scales and frequencies, but is an adequate approximation in some situations.
As we shall see, this separation provides a useful framework for interpreting exotic  correlations in the current Holometer  data because of the significant difference in scale between separation and size. 

We will not treat here the  general situation, where beamsplitter separation is comparable with or much larger than the interferometers.
That limit is more ambiguous to treat in a semiclassical way because  classical timelike trajectories such as measurement world lines, and the spacelike hypersurfaces that define classical time coordinates in any frame,  emerge from the deep quantum system, which significantly affects the projection of states onto classical  causal diamonds.  This ambiguity, and the limits of validity of our approach,  are discussed in more detail below.

 \begin{figure}
\begin{centering}
\includegraphics[height=2.4in]{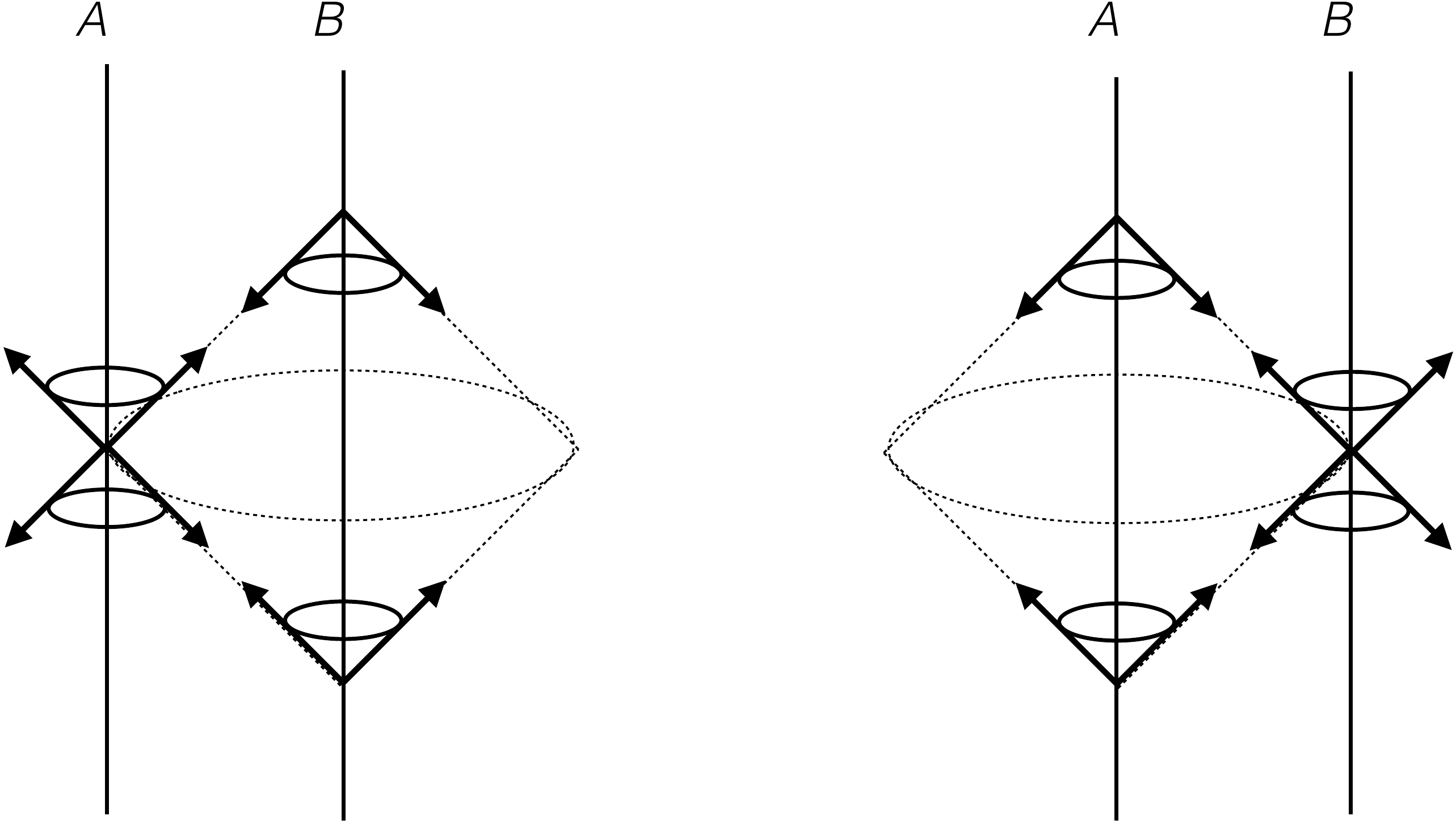}
\par\end{centering}
\protect\caption{Causal structure of entanglement information between world lines $A$ and $B$.    States of past and future null cones of an event on world line $A$ map onto the causal diamond of an interval on world line $B$, as expressed in the classical relation, Eq.\,(\ref{anti}). The complementary diagram with $A$ and $B$ reversed is shown at right. \label{pastfuture}}
\end{figure}

\subsection*{Imaginary cross spectrum from antisymmetric  entanglement of causal diamonds}

Our model for the cross correlation is based on  principles similar to those used for the autocorrelation: 
Measurements, including
correlations of measurements on different world lines,  happen on causal diamonds. 
A measurement on a causal diamond includes past and future null cone states at two events separated by some duration in proper time.
We posit that cross correlations  entangle  states on the two world lines such that the cross correlated part of the $B$ time series has  an antisymmetric relationship between the future and past null cones that connect it with the $A$ time series:
\begin{equation}\label{anti}
2\Delta_{B|A}(t) = \Delta_{A}(t+ R/c) -    \Delta_{A} (t- R/c).\vspace{0.2em}
\end{equation}
This  relationship  expresses the  principle that spacelike  relationships are represented by quantum states that live on causal diamonds (see Fig.\,\ref{pastfuture}). Planck scale information is  correlated between null separated events, at  times on $A$ and $B$ separated by $\pm R/c$.   As written here, the two terms represent the correlated information on null cones propagating into the future and past from  an event on $B$, which map onto a causal diamond for an interval on $A$  extending over a proper time  $\pm R/c$.
 
This relationship between $A$ and $B$ twists describes their noninertial relative displacements to each other. The twist of a future light cone gets ``undone'' when the return light cone of the causal diamond comes back to the same world line.
 
An equivalent  relation  must also hold for $A$ and $B$ reversed.  The choice of $A|B$ or $B|A$ represents a choice of which one ``owns'' the measurement---  which one is the axis of the causal diamond, and which one is on the surface. Physically, it represents a choice of convention for propagation direction of information along the separation axis.
The antisymmetric relationship preserves consistent parity in reflexive rotational ``displacements'' of 4-position.

Eq.\,(\ref{anti}) is classical shorthand for the the spooky correlation in a measurement of entangled states.  An entry in the $A$ time series at each proper time $t$  correlates with  values of  the $B$ time series on its future and past null cones.  The correlation expresses a concrete   relationship  arising from \textit{pure entanglement information} between world lines: the spooky correlations manifest themselves only when data from two world lines are compared.

This cross correlation encodes  information about exotic transverse ``motion''  of the two beamsplitters relative to each other--- the fluctuation of their positions from classical  ``absolute'' space, determined by the incoming null data that determines the global inertial frame.   It requires a new kind of nonlocal superposition of geometrical states, associated with emergence, that exactly preserves causal relationships.

 The quantum underpinnings of Eq.\,(\ref{anti}) should ultimately follow from the Planck scale quantum system from which space and time emerge.  
A simple quantum system described in the Appendix  (Eq.\,\ref{plusorminus})  displays a correlation of this form, based on the entangled   null cone states that comprise a causal diamond. That system is based on a set of binary relationships that do not depend on notions of space, time, or locality.  Emergent time and emergent directions inherit the antisymmetric behavior of that system.
 As discussed in the Appendix (Eq.\,\ref{plusorminus}), it can be motivated by a similar time antisymmetry  associated with antipodal directionality  in a model of particle states around an eternal black hole\,\cite{Hooft:2016cpw}. 
 For now, we treat Eq.\,(\ref{anti}) as an  {\it ansatz} motivated by the relational entanglement hypothesis, and evaluate its consequences.


 Transforming Eq.\,(\ref{anti}) into the frequency domain,  the cross spectra of  $A$ and  $B$ null cone states are related by
\begin{equation}\label{eqn:crossimag2}
2\,\tilde{\Delta}_{B|A} (f)\, =\,  \tilde{\Delta}_{A}\,  e^{+ i 2\pi f R/c}  - \,   \tilde{\Delta}_{A}\,   e^{- i 2\pi f R/c}
= \, 2  i \,  \sin (2\pi f R/c )\ \,    \tilde{\Delta}_{A}
\end{equation}
Thus,  antisymmetry in the time domain leads in the frequency domain 
to an imaginary cross spectrum:\vspace{0.1em}
\begin{equation}\label{eqn:crossimag}
\tilde{\Delta}_{B|A}  \, \tilde{\Delta}_{A|B}^* = \,   i \,  \sin (2\pi f R/c )\ \,    \tilde{\Delta}_{A}  \tilde{\Delta}_{A|B}^* \approx \,  i   \,   t_P\,  \sin (2\pi f R/c )\ \vspace{0.1em}
\end{equation}
where the last  estimate assumes  a Planck power spectral density for $\smash{\tilde{\Delta}_{A}  \tilde{\Delta}_{A|B}^* \approx t_P}$. The cross spectrum displays  {\it    the distinctive spectral signature of pure entanglement information:   the  real part of each time series projects exactly onto the imaginary part of the other, and  vice versa, with a spectrum determined by their separation}. This phase information is lost in the autospectra (the square moduli), but shows up in the cross spectrum, with the product between the signals subject to a triangle inequality.

 \subsection*{Matched-segment toy model of interferometer cross response}


In the spirit of the toy model of autoresponse, we start with a rough estimate of  cross  response
for a ``matched-segment'' configuration where $A$ and $B$ have  parallel, exactly transverse arm segments, much longer than the separation $R$.

\begin{figure}
\begin{centering}
\includegraphics[height=3.05in]{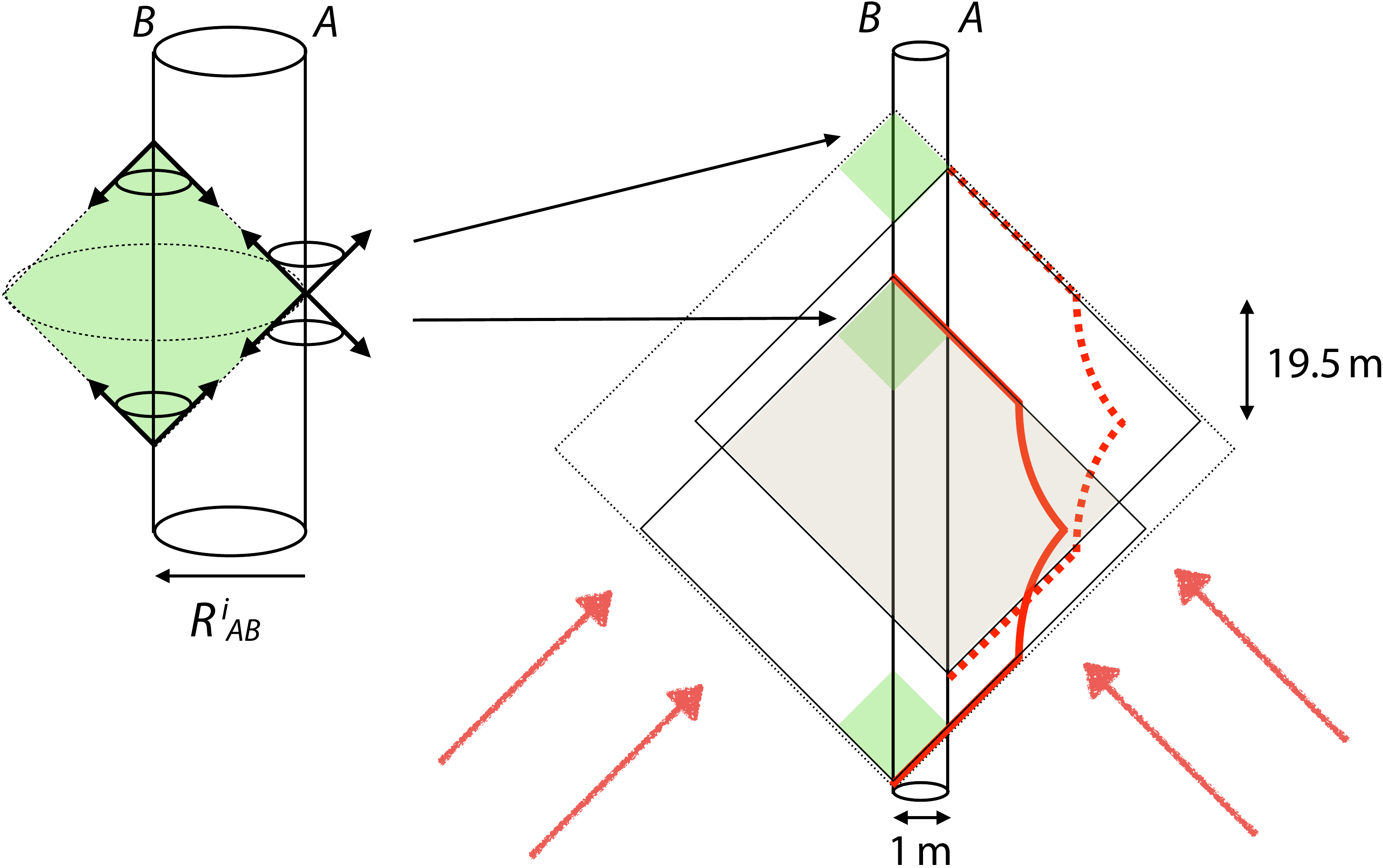}
\par\end{centering}
\protect\caption{Space-time diagram of  causal diamonds  entering into the  cross spectrum measurement (Eq.\,\ref{crossspectrum}), in laboratory time and radial coordinates. The radial components for light paths of the bent arms in two interferometers are shown for measurement of correlation at a particular time lag in $B$'s proper time, with the larger causal diamond encompassed by the cross correlation measurement that includes their separation.  The smaller causal diamond enlarged at the upper left and in Fig.\,\ref{pastfuture} represents the origin of the ``holographic noise''  from the relative rotation of two inertial frames, determined by the entanglement of the two measurement world lines, as represented in Eqs.~(\ref{anti}) and (\ref{spacecross}).  The relative ``motions'' of the two world lines are defined by their reflexive displacements relative to null trajectories in the  direction $\smash{r_{AB}^i= R^i_{AB}/|R|}$.  Incoming arrows represent components of environmental  null states along $r_{AB}^i$  that determine the relationship of the local and global inertial frames, as explained in the Appendix. The figure is not drawn to scale, but the most important scales are labeled: the $R=0.90$\,m beamsplitter separation that controls the entanglement via the antisymmetric cross-correlation scale of proper times, and the $L=19.5$\,m transverse arm lengths that set the scale of the response functions.\label{entangled}}
\end{figure}

The  cross response  exhibits corrrelations not present in the autoresponse.   Cross correlations in the differential  phase of interfering light is affected by exotic entanglement of frequencies of  photon states in the two instruments. States of light in interferometer cavities are spatially delocalized, so  
the causal diamonds for this effect on the cross spectrum include a larger time interval than the cross projections of the $\Delta$'s:  
 instead of  Eq.\,(\ref{anti}), the relevant  time interval  is the time $L_\perp/c$ between reflections of light off of mirrors in  transverse segments (see Fig.\,\ref{entangled}). 
 
For now we adopt a  simple approximation for this contribution to $\smash{ \tilde{W}_{B|A}  \tilde{W}_{A|B}^*}$,  formulated in the frequency domain. The classical picture  as a cross response to an impulse is suspect at high frequencies $\approx (R/c)^{-1}$, where the exotic entanglement affects emergent time. On the other hand, in the approximation where  $A$ and $B$ are close together ($L_\perp/c >> R/c$),  their arms  should share approximately  the same cross ``impulse'' for geometrical responses at low frequencies.  If so, the cross response at low frequency should be real, since the geometrical state vectors are in phase.

Thus, for identical, parallel, nearly-coincident transverse  segments of $A$ and $B$ of length $L_\perp$, we posit  a cross spectrum with the same sinusoidal frequency dependence as in Eq.\,(\ref{eqn:crossimag2}), but scaled to the larger causal diamond, and with a signum  added to yield a symmetric real function of frequency.
This allows us to write the  total cross response factor in Eq.\,(\ref{crossspectrum}) as
 \begin{equation}\label{crossresponse}
\tilde{W}_{B|A}  \,  \tilde{W}^*_{A|B}\, \approx  \ {\rm sgn}(f) \, \sin (2 \pi f L_\perp / 2c)
 \ \ [\hspace{.06em}\tilde{W}_{A|B}  \,\tilde{W}^*_{A|B}\hspace{.06em}]_{3D} \ . \vspace{0.2em}
\end{equation}
This separation of the 3D spatial part from the time/frequency part is approximately valid for the Holometer layout because of the separation of frequencies associated with arm lengths $L$ and instrument separation $R$. 

As explained in the Appendix, the last expression in Eq.\,(\ref{crossresponse}) in square brackets can be interpreted as a 3D projection of the $A$ autoresponse, projected onto a displacement arriving from the $B\hspace{.04em}A\hspace{.04em}$ direction. For aligned $A$ and $B$ segments with asymmetric arms (like the Holometer), it can be understood in a way similar to the $A$  autoresponse, so we adopt a toy boxcar segment model as before:\vspace{0.1em}
\begin{equation}\label{toypower2}
 [\hspace{.06em}\tilde{W}_{A|B}  \, \tilde{W}^*_{A|B}\hspace{.06em}]_{3D}\hspace{.06em} \,\approx \,\hspace{.06em} 2\hspace{.06em}(L_\perp/c)^2 \ {\rm sinc}^2   (\pi f L_\perp/ 2c)\, . \vspace{0.1em}  
\end{equation}
Putting these expressions together, the approximate matched-segment cross response spectrum is:\vspace{0.1em}
\begin{equation}\label{crossresponsetoy}
\tilde{W}_{B|A}  \,  \tilde{W}^*_{A|B} \,\approx \  {\rm sgn}(f)\,  \sin (2 \pi f L_\perp / 2c)\ \ 
 2\hspace{.06em}(L_\perp/c)^2 \ {\rm sinc}^2  (\pi f L_\perp/ 2c)\, . \vspace{0.1em}
\end{equation}
We have used in the above equations a frequency scale based on $\frac{1}{2}L_\perp$ instead of $L_\perp$. These seemingly ad hoc choices--- the prefactor on segment scales and the addition of a signum--- are explained by a physical picture of cross correlations motivated in the next section.

The spatial dependence of the cross response can be generalized covariantly to three spatial dimensions by  spatial projections of  response tensors with  frequency-independent unit vectors, as displayed below in Eq.\,(\ref{spacecross}). It is therefore always smaller than the auto response, as required for consistency.

\subsection*{Toy model of the cross spectrum}

Putting these models of $\smash{\tilde{\Delta}_{B|A}  \, \tilde{\Delta}_{A|B}^*}$ and $\smash{\tilde{W}_{B|A}  \tilde{W}_{A|B}^*}$  together, a rough but complete model of the  measured cross spectrum is:\vspace{0.1em}
\begin{equation}\\ \label{toy2}
\tilde {S}_{BA} (f) \, \approx\, i\,t_P  \ \sin (2\pi f R/c ) \ \   {\rm sgn}(f)\, \sin (2 \pi f L_\perp / 2c)\ \  [\hspace{.06em}2\hspace{.06em} (L_\perp / c)^2 \ {\rm sinc}^2   (\pi f L_\perp / 2c)\hspace{.06em}] \,  . \vspace{0.2em}
\end{equation}
This model has been constructed to respect  nontrivial  constraints consistent with expected symmetries of the system.
The cross response (Eq.\,\ref{crossresponsetoy}) is a real and even function in frequency, and an even function in time, as we argued  should be a good approximation for this setup. The cross  spectrum   (Eq.\,\ref{toy2}) is  an odd function of frequency, as  required for a purely imaginary signal that  arises from real-valued time stream data. 
It  has  zero frequency derivative at the origin,  which arises from a zero total integral in the time domain cross correlation, and thus preserves the exact antisymmetry that we have posited for the contributions from past and future null cones.


Although not entirely realistic,  this model for the cross spectrum is useful and illuminating. It allows us to trace the effect of several ways  exotic correlations in space-time can affect  a measurable signal correlation.  In this form, it has no free parameters, but as discussed below it can be generalized in ways that allow experimental tests of its assumptions.    It  shows explicitly how the shape of the  spectrum,  including the overall normalization, depends on the scales of the apparatus, $R$ and $L_\perp$. 
Most significantly, it shows  how  a purely imaginary broad band cross spectrum at low frequency arises from pure entanglement information. We know of no conventional source of correlation with this distinctive signature. Indeed, in principle, it may have no physical counterpart in a classical picture, where Green's functions apply within a local framework and relate real and imaginary parts by causality.

\section*{Realistic Models of Cross Spectra}

The model just given provides a rough guide to the expected behavior, but the  detailed forms of the response and  cross response tensors 
  are not fixed uniquely by known and well tested physical principles.
Moreover, it is well known that to analyze a system in quantum mechanics, a full account must be included of all degrees of  freedom that enter into a measurement.  As can be seen in Fig.\,\ref{entangled}, the whole measurement of a cross spectrum involves  a coherent state of a large causal diamond that encompasses the time interval of a signal measurement in both interferometers. Time and radius are not  separable in the underlying geometrical quantum system, so the cross response in general depends on entanglements of radial and transverse states that are not yet included in our semiclassical approach.

Although we cannot yet offer  a unique general prediction for a cross spectrum,  we  propose here a set of approximations for the cross spectrum that are appropriate for the  Holometer layout, which correlates signals of two similar devices much closer than their size.
This allows construction of  concrete models of cross spectra based only on known scales of the apparatus, and makes predictions for modified versions of the apparatus that allow tests of the exotic origin.

\subsection*{Cross response of arm segments from time projection}


While the matched-segment toy model (Eq.\,\ref{toy2}) was motivated in the frequency domain, in the context of the whole system (Fig.\,\ref{entangled}), the time and spatial separations between measurements shown together allow us to better conceptualize the combined toy model cross response in the time domain. This helps us naturally generalize the toy model to cross-segment correlations, such that we can take a sum of these components to construct a segmented model of the cross-response between entire $A$, $B$ interferometer paths in the following section.

\begin{figure}
\begin{minipage}[b]{.52\textwidth}
\begin{centering}
\includegraphics[height=8.1in]{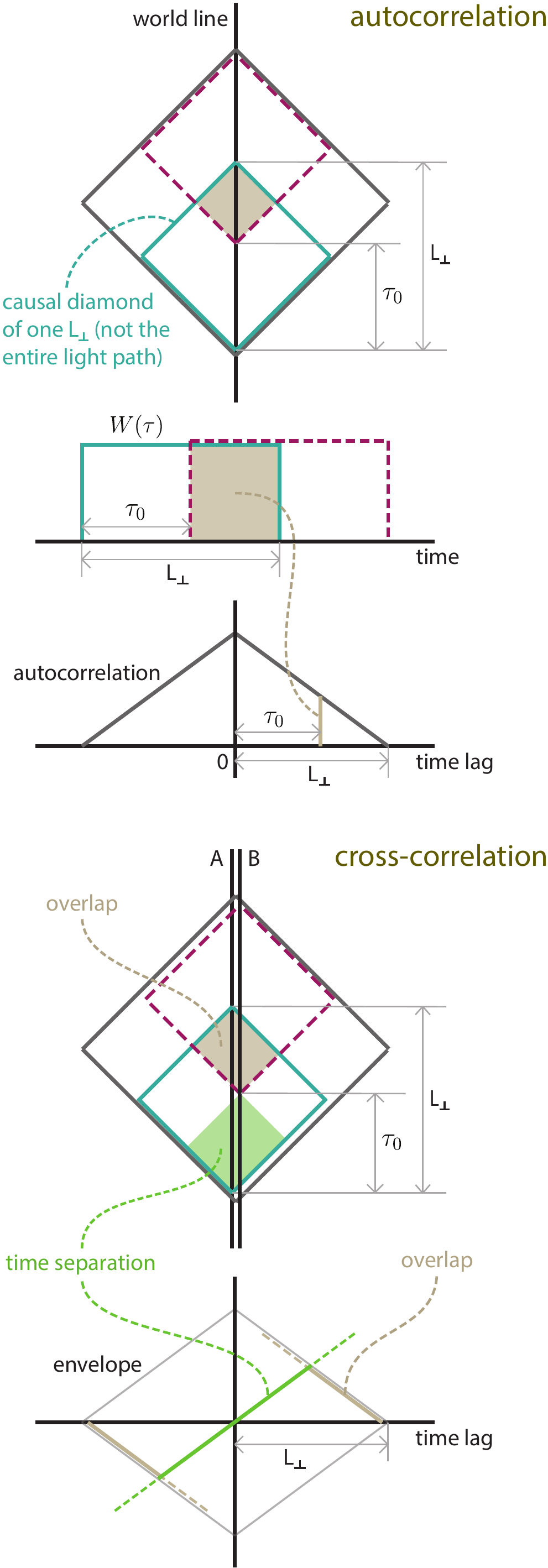}
\par\end{centering}
\protect\caption{A model of cross-correlation based on relational principles. The envelope is similar to the autocorrelation, based on the overlapping region within the causal boundaries of both A and B measurements. However, unlike the autocorrelation, the cross-correlation increases linearly as the proper time separation along the world lines increases from zero, just like it does for radial spatial separations between the world lines in Fig.\,\ref{entangled}.\label{autotocross}}
\end{minipage}
\hfill
\begin{minipage}[b]{.44\textwidth}
\begin{centering}
\raisebox{\dimexpr-\height+0.84\textheight\relax}{\includegraphics[height=7.5in]{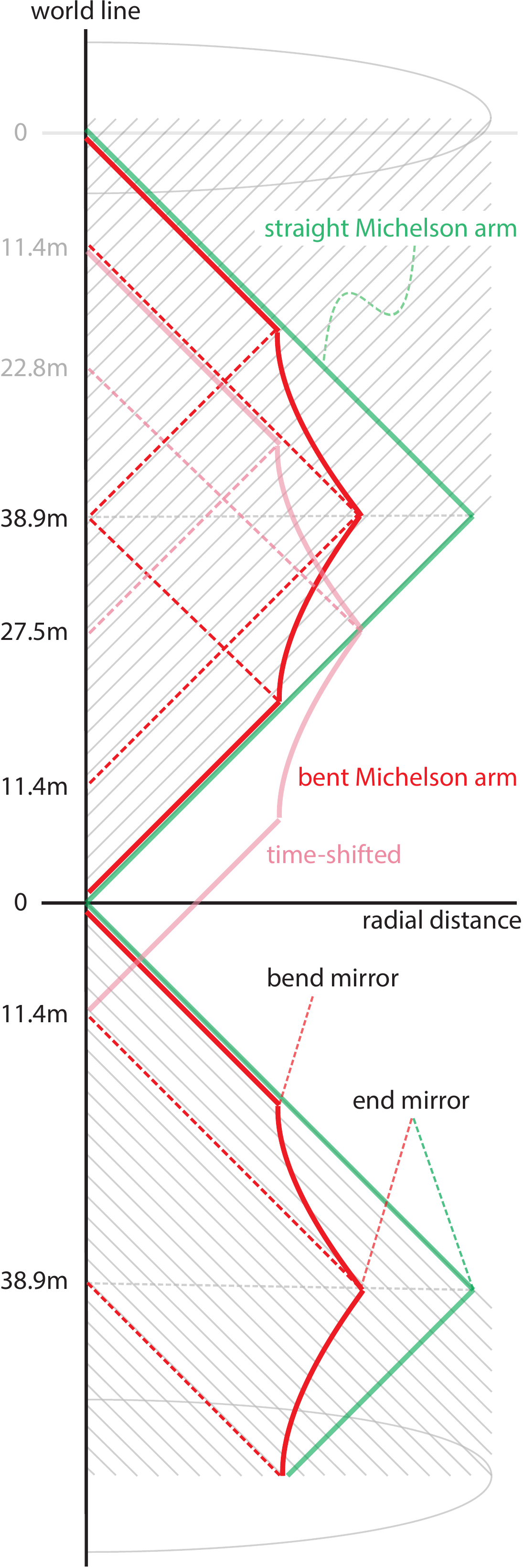}}
\par\end{centering}
\protect\caption{Space-time diagram of light trajectories, with projections and correlations along past and future null cones. Signals are sent out and measured at the world line, and can involve both directions in proper time. An illustrative example for time lag $\approx 11.4\rm{m}$ $c^{-1}$ is shown. \\ \\ \label{lightconecorr}}
\end{minipage}
\end{figure}

The ``boxcar'' autoresponse (Eq.\,\ref{toypower}) can be visualized as an effect of time lag between two signal streams: as the time lag increases, their causal diamonds   overlap less (see top panel of Fig.\,\ref{autotocross}). Each causal boundary defines the information accessible to the measurement, so the correlation is limited by the size of the overlapping region. In this model, the correlation is maximized when the proper time separation along the world line is zero.

However, in Fig.\,\ref{entangled}, we see another effect for the cross correlation $\Delta_{A|B}$ between two spatially separated world lines. The same causal envelope is present: just like in the triangular autocorrelation, as the radial separation $R$ increases to a distance comparable to the apparatus size, the overlapping region between the two causal diamonds decreases linearly, limiting the quantum geometrical information that can show up in the cross-correlation. But  a different scaling  appears at small values of $R$: the small shaded causal diamond that represents pure entanglement information in relational space-time--- the relative measurement between two spatially separated world lines--- scales linearly with $R$, and increases with the radial separation.

The universality of the speed of light defines a nonlocal invariant relationship between events: causal structure identifies radial distances with proper time intervals on each world line. If a cross-measurement of relative ``motion'' approaches zero as the \textit{spatial} separation $R$ between two observers goes to zero, then it must also become zero as the \textit{time} separation between two measurements reduces to zero.
The cross-correlation should thus change in a similar way with respect to these two distances: separations in proper time \textit{along} the world lines, and spatial separations in radial distance \textit{between} the world lines. So even considering just the $W$ time response functions without the $\Delta_{A|B}$ world line correlations, the limit of $\smash{\tilde{W}_{B|A}  \tilde{W}_{A|B}^*}$ as $R \to 0$ (or rather, the analytic extension of its values above the beam waist size) does not  represent the same physical correlation as $\smash{|\tilde{W}(f)|^2}$. The former is a cross response between two measurements with different photons, and the latter is the square modulus of the response function for a single measurement from the self-interference of each photon with a single ``clock''.


Fig.\,\ref{autotocross} (bottom panel) shows these  concepts implemented in the $W_{A|B}$ cross responses. As two causal diamonds (each representing an $L_\perp$ segment on the world line) are separated in proper time away from zero lag, the \textit{differential} cross response increases linearly; however, as it hits the causal envelope, the overlapping region of accessible quantum geometric information brings the cross response linearly back down to zero at time lag $L_\perp$. Both the segment scale $L_\perp$ and zero in the time domain  correspond to zero crossings in the correlation; indeed, all the time intervals at which the light trajectory reflects off of the optics are zero crossings.   This behavior makes physical sense, as these are the points in space-time where the interactions of light occur with world lines to change direction. This heuristic picture is consistent with triangular correlation functions of characteristic time scales $\frac{1}{2}L_\perp$ instead of $L_\perp$; the cross responses are set by half the original ``boxcar'' scale since there is zero correlation between perfectly overlapping segments.

There are  significant differences between the toy cross spectrum model (Eq.\,\ref{toy2}) and the ``boxcar'' autospectrum model   (Eq.\,\ref{toypower}). The  $\sin (2 \pi f L_\perp / 2c)$ term  comes from the effect of relational cross measurements across time separation on photon states in a whole arm, while the   $\sin (2\pi f R/c)$ term represents relational cross measurements for the beamsplitter spatial separation. They both arise from the antisymmetric offsets of the base triangles  in the time domain--- by $\pm R$, or by $\pm \frac{1}{2}L_\perp$ (see Figs.~\ref{pastfuture}, \ref{entangled}, and \ref{autotocross}). Also, the ${\rm sinc}^2$ terms, which represent these base time-domain triangles, naturally have zeroes at frequency scales twice as large. The overall prefactor is kept the same as Eq.\,(\ref{toypower})  instead of taking the classical envelope in Fig.\,\ref{autotocross} literally and scaling down from $L_\perp$ to $\frac{1}{2}L_\perp$, to respect consistent normalization in the entangled time-series fluctuations, as is done in Eqs.~(\ref{anti}) and (\ref{eqn:crossimag}) as well as the Appendix.

In a  background-independent quantum  theory, laboratory time would not appear as a variable, but would remain an observable, just like space is posited to be in our models. They would emerge from a single system. In our framework, because we use invariant classical elements as building blocks, the two distances, proper time and radial separation (1+1D around an observer), are connected along null cones that define the entanglements in space-time states. Applying the antisymmetry as two separate sine terms in Eq.\,(\ref{toy2}) can be viewed as an approximation allowed  by the fact that this configuration has two physically distinct separation scales $R$ and $\frac{1}{2}L_\perp$ that are an order of magnitude apart and in two orthogonal dimensions with respect to a world line. The signum reflects the understanding that there is only one antisymmetry in the unified system.

To further illustrate how the single antisymmetry manifests as two separate sine terms with a signum, consider the case where the time-domain cross-correlation in Fig.\,\ref{autotocross} (bottom panel) are built out of two symmetric triangles instead of two antisymmetric ones---  as time symmetry is expected in typical correlation functions of conventional phenomena. The frequency-domain cross response might take the form\vspace{0.1em}
\begin{equation}
\sim \, (L_\perp / c)^2 \,{\rm sinc}^2 (\pi f L_\perp / 2c) \ \, \cos (2 \pi f L_\perp / 2c)\,, \vspace{0.1em}
\end{equation}
instead of the version in Eq.\,(\ref{crossresponsetoy}) that has a sine function with a signum. The following identity offers a hint of how the exotic form in Eq.\,(\ref{toy2}) might then arise from the effect of  exact past/future antisymmetry:\vspace{0.1em}
\begin{equation}\label{trig}
\sin (2 \pi f R/c)\ \sin (2 \pi f L_\perp/2c) \,= \,\tfrac{1}{2}\cos\, [\,2 \pi f \,(L_\perp-2R)/2c\,] - \tfrac{1}{2}\cos\, [\,2 \pi f\, (L_\perp+2R)/2c\,]\vspace{0.1em}
\end{equation}
Thus,     a  generalized version of the antisymmetric past/future null cone construction (Eq.\,(\ref{anti}) and Fig.\,\ref{pastfuture}), in a more complete theory that treats  time and space in a unified account of cross response, might  lead to a  sine for each transverse segment in measured lab frequency domain.
On one world line, the two  triangles are  brought together by $\pm R$, and on the other, they are shifted apart by $\pm R$.
The $\frac{1}{2}$ factor gives consistent normalization in our semiclassical understanding of relational cross-correlations described above (Figs.~\ref{pastfuture} and \ref{autotocross}).


\subsection*{Segmented model of the Holometer cross-spectrum}
We proceed next to construct more complete and realistic model of  the actual Holometer cross spectrum.
It is  based on a sum of  segment cross spectra of the form just discussed, but with  more careful attention to  normalization,  parity, and frequency scales for light propagating in both directions on the arms. All of the parameters in the model are connected to known dimensions of the apparatus.  If a model similar to this is shown to fit the data, it can be used to predict the behavior of an apparatus with a different layout, as long as it is not too different.

To arrive at a complete spectrum that is usable for interpretation of actual Holometer data, we  need to make some choices beyond the approximations used in  cross spectrum toy model with one $L_\perp$ scale (Eq.\,\ref{toy2}). As shown above in Figs.~\ref{impulse} and \ref{autodecomposition}, projections and correlations along null cones suggest that the physical scales of the system implemented in the model should include $L_1 =  11.4$m and $L_2 =  27.5$m, as well as  the overall causal diamond scales in the  lab frame, $L_3=38.9$m and $L_4=77.8$m.  A sum of these linear segment responses leads to the following model:

\begin{figure}
\begin{centering}
\includegraphics[width=6.5in]{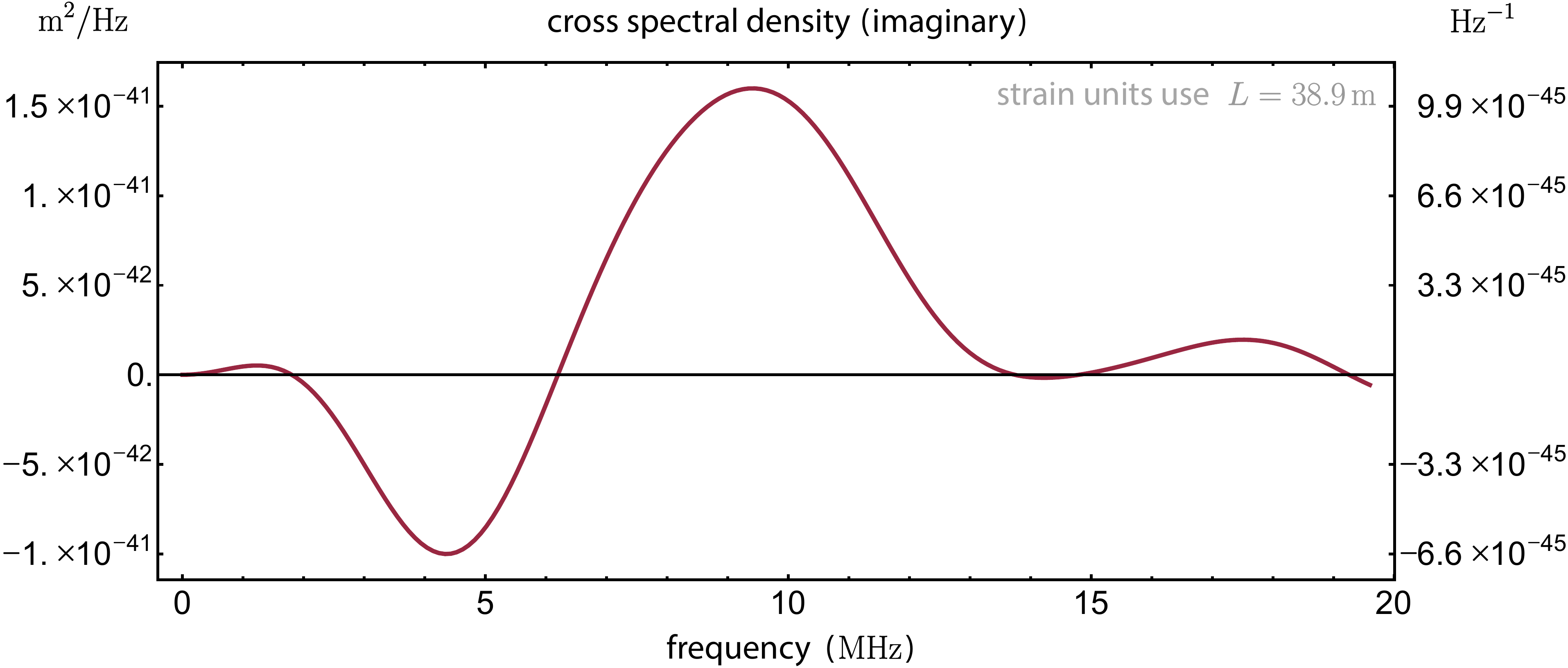}
\par\end{centering}
\protect\caption{Expected power spectral density of the signal cross spectrum between the two Holometer interferometers, estimated by the phenomenological model in Eq.\,(\ref{toyfinal}), with a Planckian normalization of $\mathscr{t}_P=t_1$ fixed by black hole entropy (see Appendix, Eq.\,\ref{norm}). The spectrum is shown in both length units and strain units, with the latter normalized to total arm length $L=38.9$m. This plot shows only the imaginary component; the real component vanishes.\label{csd}}
\end{figure}

\begin{figure}
\begin{centering}
\hspace{-0.73in}\includegraphics[width=5.97in]{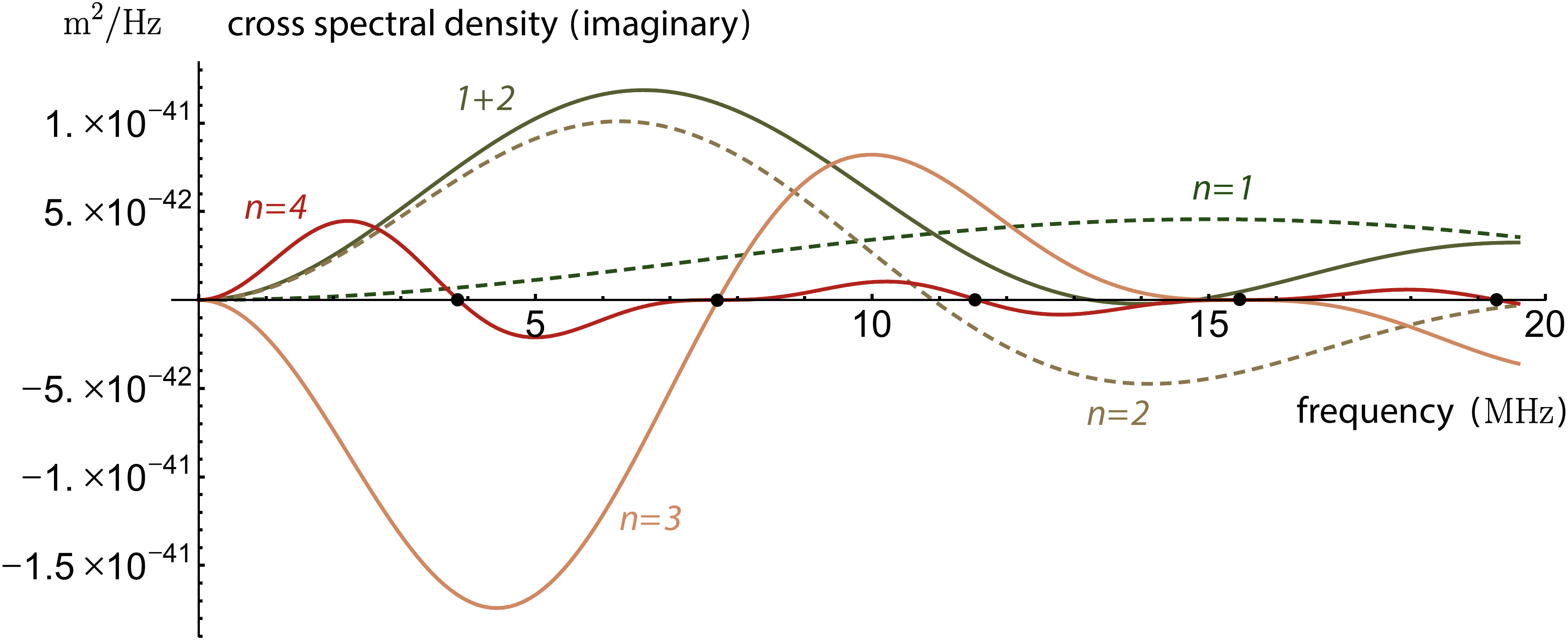}
\par\end{centering}
\protect\caption{The same expected power spectral density of the signal cross spectrum as in Fig.\,\ref{csd}, plotted component by component, as specified in Eq.\,(\ref{toynumeric}). Free spectral ranges are denoted by dots.\label{csd_comp}}
\end{figure}

\vspace*{-2em}\begin{align}\label{toyfinal}
\tilde {S}_{BA} (f) \, =\, \sum_n\, &\ 2\beta_n\ [\, 2\,(L_n / c)^2 \,{\rm sinc}^2   (\pi f L_n / 2c)\,] \ \, i\,\mathscr{t}_P \, \sin (2\pi f R/c ) \ \sin ( \pi f L_n / c)\ {\rm sgn}(f) \\ \nonumber  \,\approx\, \sum_{n=1}^4\, &\ 0.291 \,i\,\mathscr{t}_P  \,\beta_n\,  \left[\frac{38.9{\rm m}}{c}\right]^2 \frac{|\hspace{0.1em}f\hspace{0.1em}|}{15.4{\rm MHz}}\ \frac{R}{0.90{\rm m}}   \\  \label{toynumeric} & \times  \left(\frac{L_n}{19.5{\rm m}}\right)^2 {\rm sinc}^2\left[\pi\frac{f}{15.4{\rm MHz}}\frac{L_n/2}{19.5{\rm m}}\right] \ \sin \left[\pi\frac{f}{15.4{\rm MHz}}\frac{L_n}{19.5{\rm m}}\right] \\ \nonumber \\ \nonumber  {\rm where}\quad \ \ &\, L_1 = 11.4{\rm m} \ \ \beta_1 = 0.84 \qquad \qquad \, L_2 = 27.5{\rm m} \ \ \beta_2 = 0.76 \\ \nonumber &\,L_3 = 38.9{\rm m} \ \ \beta_3 = 2 \times -0.46 \quad \ \ L_4 = 77.8{\rm m} \ \ \beta_4 = 0.119 
\end{align}

\vspace*{0.8em}\noindent We have adopted here $\mathscr{t}_P$ as a normalization parameter to be tested against experimental data, expected to be on the order of Planck time $t_P$. Figs.~\ref{csd} and \ref{csd_comp} show the resulting frequency power spectrum, assuming a normalization $\mathscr{t}_P=t_1$ based on black hole entropy (see Appendix, Eq.\,\ref{norm}) that fixes the absolute normalization of the model.   The  normalizations for the individual coefficients $\beta_n$  are estimated from the calculated elements of the autocorrelation model in ref.\,\cite{Hogan:2016}, shown in Fig.\,\ref{autodecomposition}.

\begin{figure}
\begin{centering}
\includegraphics[width=6.85in]{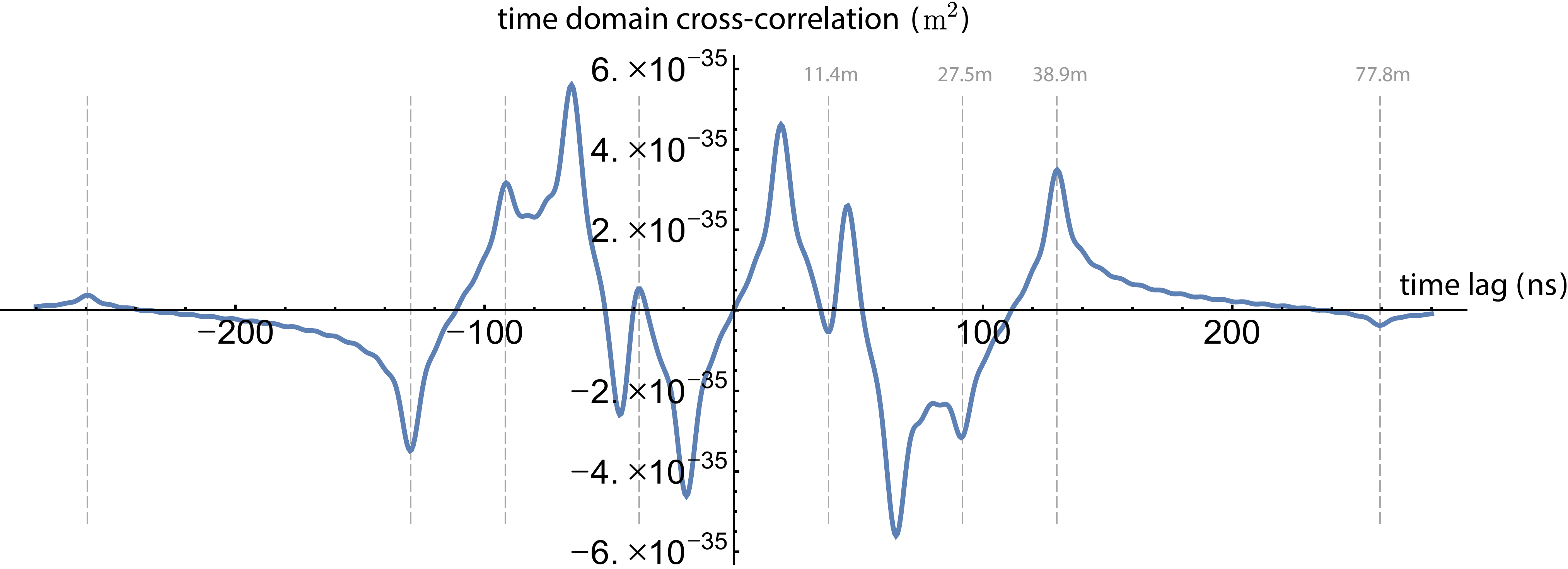}
\par\end{centering}
\protect\caption{Time domain cross correlation of  the same model as in Fig.\,\ref{csd}, showing the antisymmetry between past and future time lags.  The parity is inverted for a  cross correlation with $A$ and $B$ reversed. Principal scales of light paths in the apparatus are shown. \label{timedomain}}
\end{figure}

The $n = 1$ and $n= 2$ components, of segment scales $L_1 =  11.4$m and $L_2=  27.5$m, and their coefficients $\beta_n$, derive straightforwardly from the two corresponding positive triangular autocorrelation components in Fig.\,\ref{autodecomposition} (see caption). 
The $n=3$ component corresponds to the negative autocorrelation components in Fig.\,\ref{autodecomposition}.
The normalization for this response, $\beta_3$, is chosen by taking the average power across the slanted trapezoidal base in the negative component from Fig.\,\ref{autodecomposition},  and assuming a triangle of the same maximum variance, with an extra  prefactor of 2, since there are two combinations of such cross responses. 
The negative components from correlations between $L_1 =  11.4$m and $L_2=  27.5$m segments have been combined into a single $L_3= 38.9$m. This works because the correlation function already starts with zero variance at zero time lag (the segments are non-overlapping and barely adjacent in proper time), as it relates conjoined but disjoint light trajectories that form a round trip. This choice for $L_3$  effectively implies that the relative ``motion'' in this component is larger by roughly a factor of two, because the segments are already separated by a one-way light path at zero lag. 

A notable difference from our previous predictions for the autocorrelation (Fig.\,\ref{autodecomposition} and ref.\,\cite{Hogan:2016}) is that we now consider correlations along both future and past null cones (see Fig.\,\ref{lightconecorr}). This motivates an additional prefactor of 2 in the normalization of Eq.\,(\ref{toyfinal}). More importantly, it removes the asymmetries in the response functions that appear as slanted-trapezoid components in Fig.\,\ref{autodecomposition}, which came from outbound and inbound light path segments that were asymmetrically projected along future null cones. The simpler triangle approximation used in Fig.\,\ref{autotocross} and Eq.\,(\ref{crossresponsetoy}) is intended to capture  the computed fluctuation power and zero crossings from those calculations,  with equal weights for  past and future.

Another significant effect of the past/future null cones is the inclusion of the $n=4$ component. In ref.\,\cite{Hogan:2016}, the zero-frequency limit of the predicted autospectrum had a residual nonzero value due to the asymmetric projections along future light cones and the resulting imperfect cancellation in positive and negative correlation power over the light round trip. This residual power  cancels out when both future and past null cones are used, consistent with  an imaginary signal, and exactly odd cross functions in time and frequency. Physically, this arises because in a Michelson system, every Planck null cone impulse equally and oppositely affects light propagating in both directions on any segment.  The natural scale to use to ensure this cancellation is the  $L_4=77.8$m scale of the enveloping causal diamond for the entire cross measurement. Fig.\,\ref{lightconecorr} shows how this is the scale that accounts for both future and past null cone projections, where causal diamonds can extend towards both the future and the past from a single event on an observer's world line (with correlations possible between them). The value of $\beta_4$ is chosen from the normalizations of the other components  to make the zero-frequency power vanish in each autospectrum.

The model can consistently account for parity violation in the cross signal. In Fig.\,\ref{lightconecorr}, the future null cones are coincident with outbound light waves, and the past null cones with inbound light waves. A future null cone that originated from the $A$ beamsplitter would pass through the nested $B$ beamsplitter on its way towards the bent arms, intersecting the $B$ world line later in time than it does the $A$ world line (see Fig.\,\ref{layout}). The phase convention for Figs.~\ref{csd} and \ref{csd_comp} is such that a $+90\,$ degree phase means that $B$ is ahead of $A$ on the time axis (as in, it ``sees'' the signal later in time).  As shown in the  Appendix, the cross correlation can vanish for some layouts. 

The choices made in assembling this  specific model are motivated by a combination of basic physical principles, exact classical calculations,  mathematical constraints, intuitive guesswork,  and early hints from actual data.    The model is more physically realistic than the toy models, if still not exact, but of course it is ultimately still based on new, still-untested physical principles.
This exercise in concrete model-building is mainly intended to  illustrate how the layout of the apparatus  determines generic features of cross spectrum. Only some of  these scales are apparent from simple inspection of the cross spectra (Figs.~\ref{csd} and \ref{timedomain}); for example, the conspicuous zero-crossing feature at  roughly 6$\,\sim\,$6.5MHz  does not actually correspond directly to any scale of the apparatus, but to a  cancellation of modes of opposite sign and different zero-crossings (the free spectral ranges are at multiples of 3.85MHz; see Fig.\,\ref{csd_comp}). With high signal-to-noise data, and a variety of experimental configurations, it should be possible to test both the arbitrary choices that went into the model, as well as  the underlying principles.

\section*{Conclusion}

This basic framework for modeling the cross response will be useful  to interpret data and to test the exotic hypothesis.  In broad terms, the main  cross spectrum frequency scales are those already quantitatively predicted in our  autospectrum model\,\cite{Hogan:2016} from the known arrangement of mirrors and light paths in the apparatus. The  predicted cross spectral amplitude is many times smaller than that of the  autospectra, as a result of the small beamsplitter separation.

The most distinctive new signature of the cross spectrum for the existing instrument is its almost purely imaginary character. The antisymmetric time correlation between  two world lines   arises directly from  spooky  entanglement  on causal diamonds.  The causal structure of the  proposed states has much in common with those  already invoked in theories of entanglement entropy of holographic gravitational systems.  We currently do not know of any conventional model of cross correlated noise that can account for an imaginary, broad band cross spectrum. 

At the core, the physical content of this framework is entirely consistent with the previous autospectrum prediction\,\cite{Hogan:2016} based on statistical covariances along causal structures. All of the new features in the model presented here--- the reduced amplitude that scales with separation between measurements, the interplay of information from including both past and future null cones, the antisymmetry along separations in time and space, and the purely imaginary character reflecting the nature of entanglement in the system---  arise from the consistent application of  relational principles to cross correlations in space-time measurements.


The framework  will  be useful to inform the design of future experiments.  A straightforward  prediction is that a Holometer rescaled to a different size should display an identical exotic cross spectrum, scaled by the appropriate factors (for example, an optical table setup with $L$ scaled down by $1/10$ and $R$ scaled down by $1/4$ would produce a spectrum that is scaled up in frequency/bandwidth by a factor of 10 and scaled down in power by a factor of 40, in reference to Eq.\,\ref{toynumeric} and Fig.\,\ref{csd}). As a result, experiments can be made smaller  (and more easily reconfigurable) than the current Holometer by employing higher bandwidth detectors and data systems. Models of the exotic response can then be tested  with  experiments in a variety of configurations.  For example, in our models the cross spectrum amplitude is typically maximized at a beamsplitter separation comparable to the size of the interferometers, but it can also vanish, depending on the details of the positions and orientations. Detailed predictions for specific proposals will be provided in future work. 
 
If exotic correlations resembling these models exist at all, precision measurements of  their  structure will allow detailed study of  new quantum degrees of freedom of  space and time, and their effects  on field propagation.
Experiments could be designed to  explore  deep symmetries  associated with the origin of  causality,  quantitatively measure the finite holographic information content of space-time, and inform theories of emergent time and space, locality, inertial frames, and quantum gravity.

\begin{acknowledgments}
This work was supported by the Department of Energy at Fermilab under Contract No. DE-AC02-07CH11359. O.K. was supported by the Basic Science Research Program (Grant No. NRF-2016R1D1A1B03934333) of the National Research Foundation of Korea (NRF) funded by the Ministry of Education. We thank T. Banks, A. Chou, L. McCuller,  S. Meyer, J. Richardson, C. Stoughton,  R. Tomlin,  and K. Zurek for helpful suggestions and conversations.  We are particularly grateful to K. Zurek for bringing ref.\,\cite{Hooft:2016cpw} to our attention. 
\end{acknowledgments}


 
\newpage
 
\bibliographystyle{apsrevM}
\bibliography{cross} 
 
\newpage
 
\section*{Appendix}

\subsection*{Spatial dependence and Lorentz covariance of the cross spectrum}

The framework developed above concentrates on the time and frequency domains, based on the  invariant causal structure of a measurement defined by causal diamonds.
To  show that our framework for the cross spectrum of two such measurements  also has a consistent spatially Lorentz invariant formulation in the 3D space of two interferometers,
it is necessary to show how exotic timelike  cross correlations in two interferometer signals relate to  the spatial structure of the apparatus.

A covariant theory  must only depend on invariant directional relationships defined by the apparatus, so that the theory respects the  isotropy of space. 
In addition,  the autonomously measurable statistical properties of each interferometer should not depend on the location (or indeed the presence) of the other one, consistent with the  relational principle that  the cross correlation takes the form of pure entanglement information.

The exotic entanglement arises between the geometrical state vectors of the measurement world lines $A$ and $B$.   In  our semiclassical framework they are 
represented in the 3D rest frames of  observers  by the random variables of proper time or frequency,  $\Delta(t)_{A,B}$  or  $\tilde{\Delta}(f)_{A,B}$.   We  assume for now that $A$ and $B$ are at rest with respect to each other.  In their rest frame the   projections onto classical directions in 3-space  must be statistically homogeneous and isotropic to agree with  symmetry of the emergent flat  space-time, so they must be given by the unique antisymmetric Levi-Civita 3-tensor, $\smash{\epsilon^{ijk}}$, where $i,j,k= 1,2,3$ represent axes in a standard rectilinear coordinate system.
The projection of the state onto the proper time of the world line is given by its time series $\Delta$.   
The spatial relationship of  its local inertial frame  to the rest of the universe at all times (and separations or frequencies) can be characterized by an antisymmetric tensor,
\begin{equation}\label{framefluctuate}
\tilde{\mathcal{R}}^{ijk} =  \,  \tilde{\Delta}(f) \ \epsilon^{ijk}\,. \vspace{0.1em}
\end{equation}
Contraction with this tensor  defines  the measurement frame--- the relationship between  proper time  and spatial directions for a particular world line.

As emphasized previously, the entanglement information is delocalized in both space and time, so the ``motion'' associated with the change of direction in an exotic rotational fluctuation can be  thought of as the ``collapse'' of a quantum indeterminacy in the definition of an inertial frame.  As always in quantum mechanics, the observables of the system and its fluctuations--- and indeed, the inertial frame itself---  are most clearly defined explicitly in an operational way, in terms of a concrete measurement. In this case, the measurement involves the particular way that cross correlations between interferometers convert exotic spacelike correlations in the geometry into exotic timelike correlations between signals.

The semiclassical response function   of a single interferometer with paths in three dimensions is  in general characterized in its rest frame by a 2-index,  3D tensor function of proper time lag $W^{ij}(\tau)$ or frequency $\tilde{W}^{ij}(f)$.  Like   the  boxcar displacement of the toy model Eq.\,(\ref{boxcar}), it represents how an interferometer maps a geometrical impulse onto a timelike signal.  We have not yet developed a   physical model for $W^{ij}$ based on first principles, but we can model and constrain some of its properties.
 
The scales of the time and frequency in the response function again arise from the lengths of its arms, which are encoded in the time or frequency structure of $W^{ij}$.
 The two indices are needed to  incorporate two classical directions  associated with a response: the direction of an ``impulse,''  and an axis of rotation.
Once again, the Levi-Civita tensor  projects correlations onto classical spatial directions. In this sense, the time series defined by $\Delta$ defines a fluctuating direction in space.

The exotic signal of an  interferometer depends on its orientation and position. On the other hand, any measurable statistical property  does not. 
For example, the exotic autospectrum is:
 \begin{equation}\label{auto3D}
  \tilde{S}_{3D}(f) \,=\, |\tilde{\Delta} |^2 \, [\hspace{0.06em} \tilde{W} \tilde{W}^* \hspace{0.06em} ]_{3D} \,= \,  |\tilde{\Delta} |^2  
 \ \, [\hspace{0.06em}\epsilon^{ijk} \,  \tilde{W}^{jk} \hspace{0.06em} ]\  \  [\hspace{0.06em}\epsilon^{ilm} \,   \tilde{W}^{lm*} \hspace{0.06em}]\, . 
\end{equation}
As required by Lorentz symmetry, the observable  (Eq.\,\ref{auto3D}) is a scalar that depends only on the shape of the device, and not on its velocity,  location or absolute orientation.



In our earlier work for exotic correlations in a single interfometer, we introduced a simple model based on classical  time-of-flight, which  gives a better estimate of time projection intervals than the toy boxcar model\,\cite{Hogan:2016}. 
  In this model, the tensor factors into radial and tangential  directional dependences of the response, described by unit vectors.
Let  $\smash{d^i(\tau)}$ denote a unit vector in  the direction towards  the rest frame position on an arm at ``light travel distance'' $ c\tau$ in the future or the past from a measurement event on the beamsplitter.   Let    $\smash{p^j(\tau)}$ denote the unit vector in the direction associated with the  orientation of light itself along the path at the same position.  An interferometer  measures the difference of response in two paths, so the response in this time-of-flight model is

\begin{equation}\label{responsetoy}
W^{ij}\, =\, d^i_1\hspace{0.06em}p^j_1-d^i_2\hspace{0.06em}p^j_2\,. 
\end{equation}
A small  light path element of length $\delta l$  on one arm contributes a  response of magnitude $|\epsilon^{ijk} p^j d^i|$  for a duration equal to  $\delta \tau \approx \delta l / c$.
The response vanishes if  the arm direction is everywhere parallel to the radial direction, which was the case for the original Holometer straight-arm configuration; a nonvanishing response requires a nonzero transverse arm component $p^j$.
This  model leads to the response function  shown in Fig.\,\ref{autodecomposition}.

The time-of-flight model is in general not adequate to describe cross responses.
The  exotic cross response  of two instruments depends on both the separation and relative orientations of $A$ and $B$, and cannot be factorized into vector components, since it is a convolution over a causal diamond.
The two measurements receive correlated components of environmental geometrical information projected along the unit vector $r^i_{AB}$ that separates their beamsplitters, leading to the antisymmetrically entangled states of  future and past null cones described above (see Fig.\,\ref{entangled}).
The  cross correlation  measures the effect of exotic ``rotation'' around axes orthogonal to $\smash{r^i}$,  to which both $A$ and $B$ respond.
The directional response of the $A\hspace{.04em}B\hspace{.04em}$ cross correlation depends only  on their individual responses, and on their  alignment  with the separation vector that determines the correlated geometrical noise.
A covariant expression for the  directional dependence of the cross response with these properties can be written  as: \vspace{0.1em}
\begin{equation}\label{spacecross}
[\hspace{0.06em} \tilde{W}_{B|A}  \tilde{W}^*_{A|B}\hspace{0.06em} ]_{3D}\, =  \, \sum_{i,l} \ \  [\hspace{0.06em}\epsilon^{ijk} \,  r_{BA}^j \, \tilde{W}_{B}^{kl} \hspace{0.06em} ]^{il}\  \    [\hspace{0.06em}\epsilon^{imn}\,   r_{AB}^m \, \tilde{W}_{A}^{nl*} \hspace{0.06em}]^{il}\, ,
\end{equation}
where  an explicit sum is  shown  over common indices contributing to the two  separate signals, corresponding to correlated components of incoming geometrical information  and rotational response.  This formula shows only the factors added by 3D spatial projection, and omits the exotic  contribution discussed above from frequency entanglement of photon states (Eq.\,\ref{crossresponse}). It shows how the response  deviates from the approximation used above of perfectly aligned, transverse segments. In this approximation, the cross response of identical interferometers is purely real, so the cross spectrum (after multiplication by $\tilde{\Delta}_{B|A}\, \tilde{\Delta}_{A|B}^*$)   is purely imaginary.

This 3D projection of the cross spectrum is constructed to have no explicit dependence on  absolute position or  direction in space; for example, it is invariant for rotations around the axis defined by  $r_{BA}^j$ and $r_{AB}^m$. 
Since it is related to the auto spectrum by  projections of unit vectors,  the cross spectrum modulus squared is always less than the auto spectrum product, as required.

\begin{figure}
\begin{centering}
\includegraphics[height=2.05in]{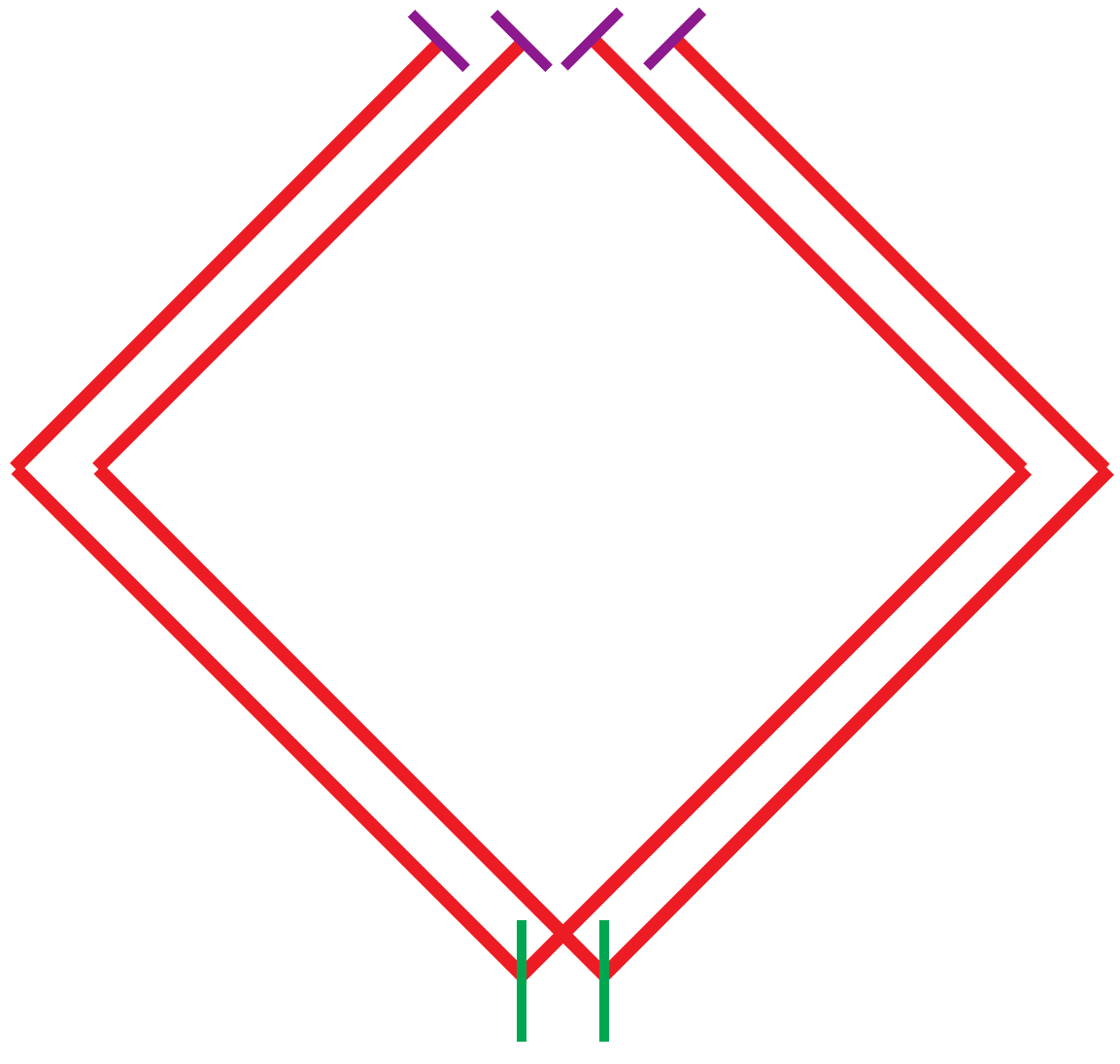}
\par\end{centering}
\protect\caption{An example of a setup with vanishing cross-correlation between two interferometer signals. For this parallel pair of Michelson interferometers with two symmetrically bent arms, and the beamsplitter separation orthogonal to their surfaces, the  spatial cross spectrum vanishes. This layout represents a symmetry where the parity cannot be determined.\label{zerocross}}\vspace*{-1.5em}
\end{figure}


The measured cross response depends  on  the paths  of light in the $A$ and $B$ interferometers, and on  their relative position and orientation in space; these macroscopic  relationships  can be modified by adjusting experimental configurations, so models of the response functions can be tested with an  experimental program.  
As one example,  Fig.\,\ref{zerocross} displays a layout configured for zero cross-response--- a symmetrical setup (with two bent arms) in which the $A\hspace{.04em}B\hspace{.04em}$ beamsplitter separation is perpendicular to the axis of symmetry. 
By contrast, the layout of the Holometer  allows for  a nonzero cross spectrum.



\subsection*{Antisymmetric cross correlations from  entanglement of null cones}

Our classical statistical models of exotic signal correlations are based on a random time series $\Delta(t)$ on each world line, and an {\it ansatz} (Eq.\,\ref{anti}) for 
an exotic antisymmetric correlation between measurements on different world lines.
A simple discrete quantum model system helps to illustrate how a counterintuitive relation of this kind can arise naturally from entanglement  of null cone states  on different world lines.

 Following  the correspondence principle, identify the classical quantity $\Delta$ with a ``world line operator''  $\hat\Delta$  that takes eigenvalues  $\Delta=\pm 1$.   Let $|+\rangle$ and  $ |-\rangle$ denote corresponding eigenstates, so that
$\hat\Delta |\pm\rangle = \pm 1 |\pm\rangle$.  In  emergent space-time, the $+$, $-$ eigenstates represent  projections of space onto proper time, and the spooky correlations of a particular series encodes the spatial relationship of a particular world line with others. Physically,  the values can be visualized  as transverse displacements (``twists'') on null cones.
Null cones originating at an event on world line $B$ intersect  world line $A$ both in the future and past.  Use $\downarrow, \uparrow$ to denote past and future directions from an event on $B$.  

Suppose that the three binary subsystems $|\pm\rangle_{A\uparrow}\hspace{0.06em},\,|\pm\rangle_{A\downarrow},\, |\pm\rangle_B\hspace{0.06em}$ are entangled. The system as a whole, which entangles null cone states of three events on the two world lines,  represents the state of a causal diamond of an interval on $A$.  It is in one of two product states:
\begin{equation}\label{plusorminus}
|+\rangle_{A\uparrow} \hspace{0.1em}|-\rangle_{A\downarrow}\hspace{0.1em} |+\rangle_B 
 \ \ \ \ {\rm or} \ \ \ \
|-\rangle_{A\uparrow} \hspace{0.1em}|+\rangle_{A\downarrow}\hspace{0.1em}|-\rangle_B \,.
\end{equation}
The quantum superposition of these two states ``collapses'' into one or the other to form the time series that represents the  measurements.
The two states display simple binary symmetries: they are invariant under exchange of $+$ and $-$, and a past/future exchange of $A$ (a time reflection) corresponds to a $+/-$ flip of the state of $B$. 
The entanglement of the three subsystems is such that there are only two independent states of the whole system: measurement (by the operator $\smash{\hat\Delta}$) of any one of the subsystems determines the state of the other two.  
Nothing in the quantum system depends on a location in  time or space, a direction,  or   a physical scale; it is neither ``local'' nor ``nonlocal.''
In our null-cone interpretation, Eq.\,(\ref{plusorminus}) describe binary  relationships between past and future time directions (corresponding to $\uparrow\hspace{0.06em},\,\downarrow$) and  left or right twist states (corresponding to  $+$, $- $),  that apply anywhere on $B$'s  null cones. 

It is straightforward to map this quantum system onto space and time using classical coordinates, if they are not  defined as part of the quantum system.
Consider   states on two world lines $A$ and $B$ that are classically at rest, so they share the same classical proper time.  
Future and past null cones of an event $B'$  on world line $B$ intersect world line $A$  at two different times. 
The times of events $B'$, $A_\uparrow'$, and  $A_\downarrow'$ are related by:
\begin{equation}\label{Btimes}
t(A_\uparrow')= t(B') + R/c   \ \   {\rm and}  \ \ t(A_\downarrow')= t(B') - R/c\,, 
\end{equation}
 where $R$ is the $A\hspace{.04em}B\hspace{.04em}$ separation. 
Application of the operator $\hat\Delta$  to Eq.\,(\ref{plusorminus})  then results in the  antisymmetric  outcome, 
\begin{equation}\label{anti2}
2\,\Delta_{B}(t) \,=\, \Delta_{A}(t+R/c) -  \Delta_{A}(t-R/c)\,, 
\end{equation}
for either of the possible  states of $B$   in the superposition, Eq.\,(\ref{plusorminus}).  
In  Eq.\,(\ref{anti}), the left side is notated as $\Delta_{B|A}$  instead of  $\Delta_{B}$; the parity changes depending on whether $A$ or $B$ is chosen to define the measurement  (the causal diamond axis in Fig.\,\ref{pastfuture}.)  Physically, this represents a choice for the direction of ``incoming'' information along the separation direction.

Measurement of  physical correlations involving both space and proper time   requires correlation of the operator $\hat\Delta(t)$  with  a clock, represented by an operator $\hat t$.
In the Holometer measurement, the beams in the straight interferometer arms, which have no response to $\Delta$, provide  reference ``clocks'' for eigenstates of beamsplitter  proper time, $\smash{\hat t_A}$ and $\smash{\hat t_B}$.   Interference  with the  bent arms creates signals that depend on  both $\hat t$ and $\hat\Delta$. 
In a simple  model where  a clock operator $\hat t$  and the  projection operator $\hat\Delta$ have a Planck scale commutator,
\begin{equation}\label{tcommute}
[\,\hat\Delta\hspace{0.06em},\hspace{0.18em}\hat t \,\hspace{0.06em}] =\, i \,t_P\,,
\end{equation}
the time series $\Delta(t)$\,---  the series of eigenvalues for the projection of $\hat \Delta$ that specifies a particular world line, whose cross correlations encode its relationship with other world lines---   displays the Planck  spectral density noise  (Eq.\,\ref{plancknoise}) assumed in our  foliation of flat space-time and in our spectral estimates.
 Eq.\,(\ref{tcommute}) can be generalized  to  covariant form with  a  4-vector displacement operator $\delta \hat x^\kappa$ and the antisymmetric 4-tensor $\epsilon^{\kappa\lambda\mu\nu}$:
 \begin{equation}\label{hatxcommute}
[\hat\Delta, \delta\hat x^\kappa]^{\lambda\mu\nu} = \,i \, t_P\,\,  \epsilon^{\kappa\lambda\mu\nu}\,.
\end{equation}
 This notation 
 incorporates the antisymmetric projections in  time and 3-space  noted previously (e.g., Eq.\,\ref{framefluctuate}). 
 


From the point of view of the measured classical time series, the time-domain antisymmetry appears  acausal,  since it involves correlations in both the future and past. In the interferometer measurement, they are actually  generated causally in the interferometer,  from spooky, spacelike correlations of geometry entangled with the states of transversely propagating light, as described by the structure of $W^{ij}$.
 The  entanglement and correlations are actually exactly causal, by construction, since the states are defined to live on null cones. 

 

\subsection*{Normalization from black hole entropy}

\begin{figure}
\begin{centering}
\includegraphics[height=3.35in]{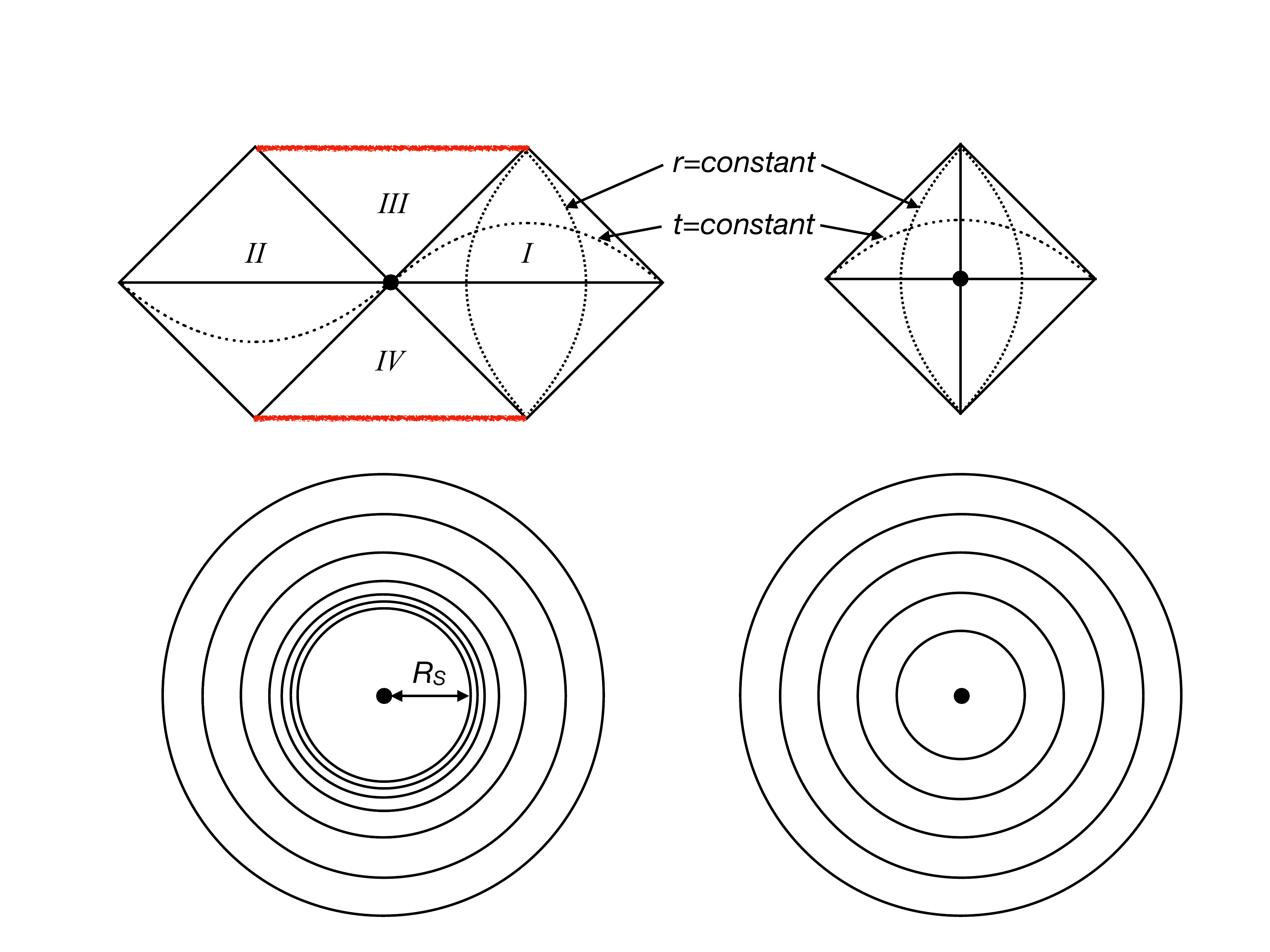}
\par\end{centering}
\protect\caption{Penrose diagrams are shown for the eternal black hole (above, left), with  regions labeled as in  ref.\,\cite{Hooft:2016cpw}, and Minkowski space (above, right). For the same solutions, the sketches below show a series of outgoing  null cones at a single time in Schwarzschild and standard coordinates in the rest frames of the black hole and the observer, respectively. If exotic correlations in the null cone states of the black hole solution have the same transverse structure as  those of our flat space-time model (that is, the foliation illustrated in Fig.\,\ref{foliation}),  the outgoing particle  states of the black hole near the horizon are entangled (and filtered) by exotic null cone correlations at the  angular scale $d\theta = c t_P/ R_S$, which leads to the correct black hole entropy (Eq.\,\ref{holeentropy}), using the normalization given by Eq.\,(\ref{norm}).  In the semiclassical model of particle states in ref.\,\cite{Hooft:2016cpw},  regions $I$ and $II$ are identified as antipodes of the same 3-space. Note the reversal of time in antipodal directions, which also appears in our formulation of  cross correlation in flat space as a form  of entanglement information (Eqs.~(\ref{anti}) and (\ref{anti2}), and Fig.\,\ref{pastfuture}). \label{hole}}
\end{figure}

Our model has one normalization parameter, the Planck scale, which can in principle be measured directly, and  determines the information content of the geometrical quantum system. It can be compared to the information content theoretically derived from thermodynamic gravity.

Start with the covariant assumption of Planck information density in proper time on any world line.
Let $t_1$ denote the  1-dimensional proper time between binary qubit states for each  of two independent axes of emergent rotation.
An interval of duration  $\tau$ corresponds to a state vector of dimension (number of states) for {\it both} axes
\begin{equation}
{\cal N}_\tau = 2^{(\tau/t_1)(\tau/t_1)}\,,
\end{equation}
and  causal diamond radius $R_0=c\tau/2$. 
This dimensionality can be compared with gravitational information, as deduced from thought experiments with black hole event horizons.
The holographic principle says that  black hole information is also the information of  causal diamonds in any space-time of the same radius and bounding area. 
 The gravitationally derived number of states for a Schwarzschild black hole of radius $R_S$ is\,\cite{Hooft:2016pmw,Hooft:2016itl,Hooft:2016cpw}
\begin{equation}\label{holeentropy}
{\cal N}_{grav} = e^{A/4}\,,
\end{equation}
where  $A=4\pi (R_S/ct_P)^2 = \pi (\tau/t_P)^2$ denotes the area in Planck units, and  {\fontsize{9.6}{0}$\smash[t]{t_P=\sqrt{\smash[b]{\hbar G/c^5}}}$} denotes the standard Planck time.
Now equate ${\cal N}_\tau = {\cal N}_{grav}$ to give 
\begin{equation}\label{norm}
t_1 = t_P \sqrt{4\ln 2/ \pi} = 0.939 \, t_P\,.
\end{equation}
This formula assumes that all the information is geometrical, and that  geometrical correlations are  responsible for entangling particle states at transverse (angular) wavenumbers that correspond to the angle subtended by a Planck length at  the horizon radius.   In actual quantum gravity, a (possibly small) fraction of the information could in principle be  carried by  other degrees of freedom, so this is a lower bound on the effective $t_1$  for geometrical degrees of freedom  that applies to   Holometer correlations. For experimental projections, we can adopt $t_P$ as a standard normalization.

Our identification of the flat-space causal diamond radius $R_0$ with the Schwarzschild radius $R_S$ in this calculation can be quantitatively motivated by using  't Hooft's  model of field states in an eternal black hole\,\cite{Hooft:2016cpw}. 
This model includes a quantum treatment of the radial component of  particles together with  radial components of gravitational ``dragging.'' In this model, 
 transverse  field operators are separable from radial ones.  The system preserves the classical radial causal structure of the geometry and the field states. For the highly symmetric case of the eternal Schwarzschild black hole, this means that the null cone foliation in the rest frame of the black hole can be matched to the  foliation in distant asymptotically flat space  adopted in our framework, so it can be used to normalize our model to gravitational entropy (see Fig.\,\ref{hole}).   The antipodes of the eternal black hole are identified with each other, but with a reversal of time, as well as an entanglement of momentum in incoming states with displacement in outgoing states, and {\it vice versa}.    The center of mass of a black hole  can have the same relationship with distant world lines in flat space as other massive bodies, with the same time antisymmetry associated with oppositely directed null trajectories that appear our models  (Eqs.~\ref{anti} and \ref{anti2}).

It was argued in ref.\,\cite{Hooft:2016cpw}  that to  agree with the holographic entropy of the black hole, it is necessary to insert a ultraviolet cutoff on the transverse partial wave components, at the angular scale subtended by a Planck length at the horizon.
Entanglement of transverse field states on null cones, as proposed here for flat space,  would provide a concrete way to   implement this filter in the emergent black hole space-time.  The angular size of a spacelike interval $ct_P$ on an outgoing null surface near the horizon (that is, at Schwarzschild coordinate $r= R_S+ \epsilon$, infinitesimally outside the event horizon)  is $d\theta=   ct_P/R_S$.  Differences in classical directions with respect to the black hole center of mass, in the asymptotically-flat rest frame, are not defined at smaller angular scale, so there is a cutoff in field state transverse wave number, corresponding to  a Planck length near the horizon,   directly from blurring of emergent directionality for each null cone. The  states of outgoing null cones, after they are far away from the horizon,  yield the same transverse correlations as the above flat-space analysis, in the proper time of distant observers at rest in the rest frame of the   black hole.
 
This derivation of normalization thus appears to be consistent with the requirement that exotic spacelike correlations of world lines in the emergent, asymptotically flat  space far from the hole should have the same structure for  black hole world lines as they do for optical elements of an interferometer.  
The system obeys a form of equivalence principle, as stated above:   {\it When interpreted correctly in terms of concrete observables,  the exotic correlations should have  the same physical effect on states of light propagating tangent to a light cone in flat space-time as they do on tangential components of field states near a black hole event horizon} (Fig.\,\ref{hole}).

\subsection*{Emergence of locality and quantum limits to the validity of the space-time approximation} 

Exotic cross correlations depend on the  relationship of world lines with each other as they emerge from a quantum system.
Our models are based on  idealized sequences on separate world lines that  represent entangled states of null cone foliations of the whole space-time.
This picture is self consistent for world lines at sufficiently large separation, but it  cannot hold all the way to the Planck scale for all world lines.
Localization of ``events''  in space and time, and  differentiation of spacelike from timelike relationships, are properties that emerge gradually on scales larger than the Planck length.  There is a duration-dependent minimum mean separation within which  world lines cannot be distinguished from one another.
 
In our relational model, most of the information on a single Planck bandwidth world line  encodes short durations and small separations.
Because of holography, its  relationships with other world lines at separation $>c\tau$, in causal diamonds  of  duration $>\tau$,   occupy a fraction $\approx t_P/\tau$ of the available entanglement information.
But if  there were more than $\tau/t_P$ other world lines within radius $\approx c\tau$,  their mutual entanglement  information would exceed the total available bandwidth for each one, so the approximation becomes inconsistent.  

This constraint imposes a minimum mean separation between world lines: in a volume of radius $>c\tau$, distinguishable world lines must have a minimum mean spatial separation 
\begin{equation}\label{Rmin}
R_{min} \approx ct_P (\tau/t_P)^{1/3}\,.
\end{equation}
That is,  Planck bandwidth world  line states can  be used to describe distinguishable, relational space time states consistent with holographic information only for larger mean separation than $R_{min}$.  Eq.\,(\ref{Rmin}) is derived in the rest frame; to use the same approximation in a boosted frame, the rest frame separation must be larger. 

At closer mean spatial separations, the classical space-time approximation is not valid. A family of world lines with smaller mean separation is substantially correlated even at the Planck frequency, so they are not distinguishable from each other; space and time have not yet distinctly ``emerged'' from the quantum system. 
In this sense, the emergence of  classical spacelike relationships depends on  the duration of a causal diamond or measurement.
The spacelike scale $R_{min}$  represents  a duration dependent separation where the world line concept itself is actually invalid.

This consistency scale is smaller than $ct_P (\tau/t_P)^{1/2}$,  the size of the accumulated rotational displacement over duration $\tau$.
Thus, the framework we have used to describe the exotic effect in terms of invariant classical concepts such as world lines and null cones is self consistent.   Our   covariant  inclusion of transverse degrees of freedom differentiates Eq.\,(\ref{Rmin}) from  similar previous limits based on Salecker-Wigner-type thought experiments about quantum limits to positional measurements\,\cite{Wigner1957,SaleckerWigner1958,DiosiLukacs1989,NgDam1994}.

At a  spacelike separation $<R_{min}$, the classical  space-time approximation of independent world lines  breaks down.
The  approximation is good to high precision in current experimental applications, but its breakdown may lead to new kinds of  connections among different sectors of physics.  For example,   
 a causal diamond duration at the Terascale ($c\tau \approx 10^{16}\ell_P$) implies a blending of world lines at   the Grand Unification length scale, $R_{min}\approx10^{5}\ell_P$.  This coincidence suggests that unification at that scale, even though much larger than the Planck length, may display significant effects of quantum gravity not accounted for  in field theory models.   
 An experimental confirmation of  nonlocal geometrical correlations could  also open up new approaches to the cosmological constant; exotic correlations in the Standard Model field vacuum could  determine the value of  its emergent gravity, as the scale of the present day cosmic horizon ($c\tau \approx 10^{61}\ell_P$) connects to a localization of $R_{min}\approx10^{20}\ell_P$ at the QCD scale \cite{Hogan:2015b,Hogan2018}.

Quantum fluctuations are widely thought to affect the classical space-time metric during cosmological inflation.  The gravitational imprint of exotic  correlations might also observably affect  quantum fluctuations during inflation, even if it occurs far below the Planck scale. The ``frozen in'' exotic quantum fluctuations of emergent gravity could modify or even dominate the effects of field degrees of freedom on large scale cosmic metric perturbations in gravitational potential. The amplitude of exotic  scalar and tensor modes would then depend on the causal diamond size, or horizon scale during inflation, rather than the parameters of a scalar field potential.  Since the exotic fluctuations  vary as a power of a slowly varying inflationary expansion rate,  they would  display a spectral tilt  similar to that predicted in slow roll inflation.


\subsection*{Response of spatially extended states of light in interferometers} 

It is worth remarking on important limitations of the classical time-of-flight model used in this manuscript to describe the propagation of light in an interferometer.  The actual response to geometrical noise depends on the structure of eigenstates of light in the interferometer.  The nonlocal correlations of  these states in both frequency and position is one rationale for the  particular exotic response term introduced in Eq.\,(\ref{crossresponse}).

Photon states in an interferometer are  delocalized. Entering states of  photons  have with a width of only about 1 kHz in the Holometer due to the power recycling cavity, so they are highly delocalized in the time domain, with correlations over about a millisecond. Single-photon states respond to exotic correlations in both arms of the interferometer throughout the round trip.  Each photon carries  information about the whole system in the frequency and phase structure of its wave packet. 

As shown in classic work by Caves and Schumaker\,\cite{1980PhRvL..45...75C,PhysRevA.31.3068,PhysRevA.31.3093}, shot noise  in an interferometer arises from  random phases of vacuum noise at all frequencies that enter the system from outside through the dark port\,\cite{Danilishin2012}.  They interfere with the correlated states inside, and  determine when photons are emitted at the dark port. 
Inside the interferometer, the emission corresponds to a transition between states of different photon occupation numbers. (A technique has been proposed to beat this shot noise limit in a holometer-like instrument, by correlating input states\cite{2013PhRvL.110u3601R}.)

In the Holometer, the photon states are  bent in the middle, and the 
 exotic cross signal is caused by rotational geometrical fluctuations.
The exotic cross correlation adds to the standard shot noise from the electromagnetic vacuum entering at the dark port:  it affects the light  over nonlocally over an interval estimated  classically by its ``light cone length'' in the detected time stream. 

The  geometrical fluctuations enter the system not just through the dark port, but from all directions on   null surfaces--- the past and future null cones of  events on the beamsplitter world lines.  
Signal correlations in exiting light are shaped by the photon wave function inside the interferometer, and its response to this geometrical noise. The familiar conjugate relations of time and frequency do not apply to these states until they are ``measured'', that is, until they emerge from the quantum part of the system as propagating states at the beamsplitter surfaces. 
These considerations help to explain  why,  for exotic correlations,  we cannot use the standard approximations for interferometry, where classical space-time response is calculated  separately from the quantum noise. 


A complete theory of emergent space-time would also have an operational definition of time, where reference clocks are needed for its measurements. The invariant framework here attempts to describe and interpret real measurements within a limited scope. Many frequencies are present in a complete characterization of the system and measurement:  the Planck frequency itself ($\smash{10^{43}}$\,Hz), the duration-dependent localization limit on emergent world lines described above ($\smash{10^{31}}$\,Hz for this apparatus), the correlation amplitude scale ($\smash{10^{25}}$\,Hz), the laser frequency  ($\smash{10^{15}}$\,Hz), the beam waist transverse size ($\smash{10^{11}}$\,Hz), and the apparatus arm length size ($\smash{10^7}$\,Hz, the experimental measurement band). The framework here only addresses measurable cross correlations in the last of these, and how they depend on the fundamental scale, the Planck frequency. The models display the lack of separability inherent in exotic correlations, but they are not a complete theory of the system. 




\subsection*{Exotic correlations in unified theories}


While our system does not require a particular consensus for a unified theory, one candidate framework might help put it into a broader context: the general theory of a holographic quantum system laid out by Banks and Fischler\,\cite{Banks:2013fri}. This framework is general enough to approximate field theory and string theory in appropriate limits. In principle it can accommodate all the holographic correlations of geometry, including the apparent delocalization of information implied by AdS duality and  black hole information paradoxes, such as  the so-called firewall problem. It naturally accommodates a thermodynamic or entropic view of gravity, in which the equations of general relativity--- the dynamics of space-time--- are a statistical description  of an emergent behavior, analogous to fluid dynamic description of molecular motions, and flat  space-time is the ground state of a quantum system.

In that theory, every causal diamond has its own Hilbert space and its own Hamiltonian. Entangled states of different world lines undergo ``fast mixing'' over intervals longer than their separation. To quote from Banks\,\cite{Banks2011} (using his notation where $N$ is the Hilbert space dimension rather than information): ``\dots the [causal] diamond operator algebras are all finite $N \times N $ matrix algebras, operating in a Hilbert space of dimension $\smash{e^{A/4}}$, where $A$ is the area in Planck units, of the holographic screen\dots\,\,The Holographic Principle tells us that both the causal structure and conformal factor of the space-time geometry are encoded as properties of the quantum operator algebra. The space-time geometry is not a fluctuating quantum variable, but, in general, only a thermodynamic/hydrodynamic property of the quantum theory."  

The Holometer experiment, in this context, is designed to capture some of the macroscopic spacelike holographic correlations before they fast-mix away, and convert them into  time-like cross correlations in measured signals. In our covariant framework, the null cone states of an interval of proper duration $\tau$ on each beamsplitter world line form a $\smash{2^{(\tau/t_P)}}$ dimensional state vector. (A unitary operator on this vector is then a $\smash{2^{(\tau/t_P)} \times 2^{(\tau/t_P)}}$ matrix.) As in the Banks-Fischler states, information is delocalized across a causal diamond: geometrical states are not independent, but entangled. Our model  for cross correlations from entangled world lines sheds some light on  how locality emerges in the Banks-Fischler theory, within a domain of validity described above.

\end{document}